# Infant FreeSurfer: An automated segmentation and surface extraction pipeline for T1-weighted neuroimaging data of infants 0-2 years

Lilla Zöllei[1], Juan Eugenio Iglesias[2], Yangming Ou[3], P. Ellen Grant[3] and Bruce Fischl[1]

[1]Laboratory for Computational Neuroimaging, Martinos Center for Biomedical Imaging, Massachusetts General Hospital
[2]Translational Imaging Group, University College London
[3]Laboratory of Fetal and Neonatal Neuroimaging, Boston Children's Hospital

**Abstract:**

The development of automated tools for brain morphometric analysis in infants has lagged significantly behind analogous tools for adults. This gap reflects the greater challenges in this domain due to: 1) a smaller-scaled region of interest, 2) increased motion corruption, 3) regional changes in geometry due to heterochronous growth, and 4) regional variations in contrast properties corresponding to ongoing myelination and other maturation processes. Nevertheless, there is a great need for automated image-processing tools to quantify differences between infant groups and other individuals, because aberrant cortical morphologic measurements (including volume, thickness, surface area, and curvature) have been associated with neuropsychiatric, neurologic, and developmental disorders in children. In this paper we present an automated segmentation and surface extraction pipeline designed to accommodate clinical MRI studies of infant brains in a population 0-2 year-olds. The algorithm relies on a single channel of T1-weighted MR images to achieve automated segmentation of cortical and subcortical brain areas, producing volumes of subcortical structures and surface models of the cerebral cortex. We evaluated the algorithm both qualitatively and quantitatively using manually labeled datasets, relevant comparator software solutions cited in the literature, and expert evaluations. The computational tools and atlases described in this paper will be distributed to the research community as part of the FreeSurfer image analysis package.

**Highlights:**

- FreeSurfer is a widely used and evolving processing suite for brain MRIs.
- Morphometric brain analysis in infants has lagged behind that of adults.
- Our novel pipeline accommodates T1-weighted brain MRIs from 0-2 year-olds.
- Its unified approach is valid across a full age range, without foregoing accuracy.
- Similar applications are largely derived from newborns only (often preterm).

**Introduction:**

Automated brain image segmentation of postnatal infant scans has received increased attention in recent years, given the unmet needs of clinical and neuroscience applications [1-3]. The age range we address in this paper is the first 2 postnatal years, during which myelination as well as brain size and shape are most rapidly changing and begin to plateau. We have chosen the term "infant" as a default descriptor, loosely referring to children during the earliest period of life.

FreeSurfer [4], a widely used processing suite for brain MRIs, has evolved for nearly 20 years. The algorithms in FreeSurfer were originally designed for and extensively tested on adult datasets [5-16], but their use in children as young as 4.5 years has also met with success [17]. Further extending the present capabilities to encompass the full postnatal period would be highly beneficial. Not only have alterations in cortical morphologic measurements (including volume, thickness, surface area, and curvature) been associated with neuropsychiatric, neurologic, and developmental disorders in children [18-21], but a greater prevalence of those surviving perinatal injuries is now apparent [22-25]. Such infants grow to adulthood, exhibiting multiple sequelae that are poorly understood. This trend intensifies the need for a unified image-processing approach applicable to a broad range of ages, as opposed to single time-point scenarios. It is important to note that a unified method does not imply a singular procedure for all participating ages but rather the harnessing of tools that automatically adapt to various input images, while yielding comparable quantitative measures as outcomes.

Most existing solutions focusing on pediatric image analysis are heavily specialized with respect to specific age ranges and imaging modalities and differ vastly in terms of resultant segmentation information. Thus, direct comparisons and evaluations remain a challenge. Strict intensity-based segmentation of these images is a difficult task, due to contrast intensity reversal in younger (vs. older) subjects (related to myelination), the relative excess of motion typically found in scans from this population, and the diminutive overall anatomy relative to voxel resolution. For these reasons, a majority of currently available solutions utilize prior information to varying degrees, which we summarize below.

*1.1 Atlases*

Many automated segmentation tools rely on the creation and usage of training datasets. These encode information on the population of interest and are often inseparable from the segmentation tools that they support. The information stored may refer to anatomic regions of interest (ROIs), average intensity, image intensity distributions, and/or or tissue probability maps. Given the multitude of challenges in infant populations, few methods rely exclusively on intensity information from input images without guidance from such sources [26, 27].

Training datasets often include manually labeled regions related to the anatomy studied. The information embedded in individuals of the set could be summarized into a single probabilistic atlas (parametric approaches) or used individually, later combining the results (multi-atlas segmentation, non-parametric approaches). In this paper, the term *atlas* refers to one member of the training dataset (MRI volume and corresponding manual labels), whereas *probabilistic atlas* entails an average volume and corresponding label probabilities. Of note, some probabilistic atlases may only include intensity information, only label probabilities (e.g., SPM [28]) or both (e.g., FreeSurfer [4]).

It is challenging to directly compare the nature and the performance of existing infant training datasets with respect to segmentation, considering the wide variability of age ranges represented, modality of images they rely upon, the number of subjects they contain or summarize, their representation, the nature of the training subjects (for example, prematurely born or full-term newborn infants), the origin of the ROI labels (manual annotation directly drawn on the training set subjects or labels projected onto the training data sets) as well as the type of information that they contain, whether it is (sub)cortical labels, average intensity values, tissue probability maps, white matter pathways, or fractional anisotropy (FA) maps. Generally, they are characterized jointly by the tools using them, which in our case are full brain segmentation solutions.

Training datasets published in the literature to date include the following: **(i)** UNC: 0-1-2 [29] (N=95, M=90 labels drawn from Automated Anatomical Labeling (AAL) map [30], K=1 atlas); **(ii)** the UNC cortical [31] (N=35, K=7 [time points]); **(iii)** the Imperial Pediatric Atlas [32] (N=33, M=83, K=33; ROIs derived from 30 manually labeled adults); **(iv)** the Imperial Neonatal Atlas [33] (N=153, M=6, K=1; average intensity and tissue probability maps and labels extracted from three neonatal reference subjects); **(v)** the Imperial Spatio-Temporal Atlas [34] (N=204, M=6, K=17; relying upon Imperial Neonatal Atlas); **(vi)** the Imperial ALBERTs [35] (N=20, M=50, K=20); **(vii)** the USC [36] (N=105 + 49, K=13); **(viii)** the INSERM atlas [37] (N=20, K=1; average T2-weighted intensity); **(ix)** Akiyama atlas [38] (N=60, M=116; mapped AAL labels; K=1); **(x)** Singapore [39] (N=112 and 32, K=2; average intensity, FA, and DTI color map); **(xi)** the JHU: neonate atlas [40] (K=1; labeled from a manually annotated single subject; M=122 structures based on diffusion-based imaging and fiber pathways); **(xii)** M-CRIB [3] (N=10, M=100; cortical and subcortical labels matching the Desikan-Killiany parcellation [41]); and **(xiii)** our Infant FreeSurfer atlases [42] (N=26, M=32+14, K=26).

Considering the availability of clinical data and clinical interest in prematurity, many neonatal training datasets are built on images of prematurely born subjects obtained at term-equivalent ages [32-35]. This is important to note, because reliance on applications rooted in prior information may introduce bias. For a more comprehensive summary of the above information, see Appendix Table 1.

*1.2 Segmentation tools*

A majority of existing postnatal infant segmentation tools are restricted to analysis of newborns [26, 27, 43-50], often focusing on or explicitly accommodating preterm subjects [26, 27, 43, 49]. Other algorithmic solutions have been designed for discrete age points within the first postnatal year (0, 3, 6, 9, 12 months) [45] or for 2 year-olds [32, 51]. Although primarily introduced to evaluate single-time point acquisitions, some require or accommodate access to longitudinal intra-subject imaging series, thus facilitating segmentation of more challenging younger-aged subjects [46, 47].

Given the relatively higher contrast between cerebral tissues in the immature brain [52], most tools currently used at the newborn stage rely on T2-weighted MR input images, either in part or entirely. Some require only a single modality [32, 43, 48, 49], whereas others use multiple channels [27, 44-47, 50]. We are aware of only one pipeline that accommodates a single T1-weighted volume for segmentation of newborns or 2 year-olds [32, 43]. In our view, relying on T1-weighted MPRAGE scans is preferable, as they are the only volumetric sequence acquired at all age groups in most clinical protocols. Indeed, these can be obtained at 1-mm isotropic resolution during a reasonable scan time, due to the ability to accelerate in two planes. In contrast, volumetric T2-weighted images are not used in infants due to poor contrast. Typically, only 2D T2-weighted MRI sequences are acquired with high in-plane resolution, but at 2.5- to 4-mm thickness, which is insufficient to resolve cortical folds.

The labels produced by current segmentation solutions also vary to a large extent. Most pipelines are aimed at tissue segmentation, i.e. labeling cortical gray matter [GM] or white matter [WM], often reflecting myelinated and unmyelinated areas, and cerebrospinal fluid [CSF]) [26, 27, 44-48, 50]. Likewise, brainstem and cerebellum [27, 48] are often labeled. Various sets of cortical and subcortical regions, including those matching the AAL [30] atlas description [29, 38, 46, 47], regions of interests defined by [30, 53] in in [32, 35], and cortical and subcortical information matching FreeSurfer labels [3] have also been used, and myelinated vs unmyelinated WM labels recovered [50].

In some segmentation frameworks, age-specific infant atlases (for subjects ≤2 years of age) are used as guidance [32, 43-45, 48-50]; but in others, segmentation labels are extrapolated from adult-based atlases. For example, the AAL atlas, derived from the anatomical parcellation of a spatially normalized single adult subject, is often invoked. In [46, 47], high-resolution T1 volumes are transferred from older pediatric subjects (2 year-olds) to segment newborn acquisitions. Alternatively, in the Imperial:Pediatric tool [32], a set of manually labeled adult acquisitions are used as prior information (30 adults, 83 ROIs). In one application [45], the authors use a semi-automatically populated infant population (<1 year of age) for this purpose.

A more comprehensive summary of above information is found in Appendix Table 2.

*1.3 Surface extraction*

Currently, only a few infant image-processing packages generate cortical surface models. Some authors [52] have segmented and reconstructed surfaces in 3D using image post-processing tools adapted from sequences developed for brains of adults [54] and fetuses [55]. Specifically, NEOCIVET [56] was introduced as a modification of the adult image-processing pipeline CIVET [57, 58] and made applicable to preterm data. A surface-based probabilistic atlas of human cortical structure from 12 healthy term born infants has been created [59], and a 4D high-definition cortical surface atlas of infants [31] has been computed using a topology-preserving deformable surface method [60, 61]. Two datasets released by The Developing Human Connectome Project (dHCP) [62] have involved minimal pipeline processing [63], relying on a set of tools (i.e. a deformable model [64], a spherical projection [65], and the FreeSurfer white matter inflation tool [66]) for approximating surfaces to independently computed cortical GM and WM segmentation labels [67].

*1.4 Contribution*

Our proposed tool is an automated segmentation and surface extraction pipeline designed to accommodate clinical infant T1-weighted brain MRIs from a population of 0-2 year-olds. The algorithm only requires a single channel MRI volume and produces automated segmentations of cortical and subcortical areas of the brain, including volumes and surfaces. The segmentation procedure adapts to the detected (or developmental) age of input data, allowing use of subsets within a manually labeled database that are optimal for each age range. This results in a unified procedure that can be applied across the full age range of interest, without sacrificing accuracy. Although the current pipeline is designed for T1-weighted images, equipped with appropriate training datasets, it would be straightforward to extend it to accommodate T2-weighted image volumes or multi-channel datasets.

**Materials and methods:**

Our pipeline is a multi-stage process closely following the adult-oriented reconstruction pipeline of FreeSurfer [5, 68, 69]. The outputs generated are consistent with its reconstruction stream, facilitating consistency in future longitudinal studies. Figure 1 demonstrates the major image processing steps in the standard FreeSurfer reconall pipeline, where red boxes indicate the ones that are different and were specifically introduced for an infant population. In this section, we focus on these particular algorithmic components.

*2.1 Skullstripping*

Extraction of brain tissue and exclusion of the skull and extra-meningeal tissue from input images are crucial early steps of any neuroimaging pipeline. Infant MRIs show large inter-subject variability, due to rapid and heterogeneous brain development. There is also a less conspicuous gap between the cerebral cortex and the skull, and lower contrast is encountered among assorted cerebral tissues. Many automated skullstrippers were primarily designed for adults [70-75] and underperform on newborn datasets. Furthermore, the existing and publically available tools specifically introduced for infants [51, 76-78] do not consistently accommodate our T1-weighted clinical datasets.

We used our novel double-consensus skullstripping approach to identify brain regions [79], applying a modified version of a tool developed by Doshi et al. [75]. This solution involves multiple pediatric datasets (12 constructed from the NIH-PD database [80] and 3 additional, randomly selected samples from secondary pediatric training datasets [32]) and multiple skullstrippers (BET [70], BSE [81], 3dSkullStrip [AFNI] [82], HWA [74], and ROBEX [83]). The first consensus is initialized via multi-atlas skullstripping solutions, and the second fuses carefully selected and optimized masks into a final result. Importantly, our method selects parameters that are optimal for each subject (not the entire dataset!). Further details are provided elsewhere [79].

*2.2 Volumetric segmentation*

We designed a multi-atlas label fusion segmentation framework [84] where ground-truth information from our labeled training data could be used for the segmentation of new infant brain images. Our solution was inspired by [85] and [86], which is in turn an MRI-contrast adaptive version of the Bayesian multi-atlas algorithm proposed by Sabuncu et al [87]. The method relies on a generative model of imaging data, which is represented in Figure 2, and uses Bayesian inference to compute the most likely segmentation. In short, this method assumes that a number of atlases have been registered to the scan to segment, provided $N$ different candidate label maps $L_n, n = 1, ..., N$. A discrete membership field $M(x) \in \{1, ..., N\}$, which is assumed to be a sample of a Markov Random Field parameterized by $\beta$ (and thus smooth), subsequently indexes from which atlas the segmentation of the target scan $L$ has been generated (Eq 1). Given the membership at a voxel $x$, the segmentation $L(x)$ is assumed to be a sample of a logOdds model defined on the distance transform of $L_{M(x)}$ (Eq 2). This segmentation is generated independently at each voxel. Given the segmentation $L$ of the test scan, its intensities I are assumed to be independent samples of Gaussian distributions parameterized by label-dependent means and variances $(\mu_L, \sigma_L^2)$, further corrupted by a multiplicative bias field. This bias field is assumed to be non-negative and smooth. Therefore, we model it as the exponential of a linear combination of smooth basis functions $\psi_p$ (Eq 4).

$$M \sim \frac{1}{Z(\beta)} \prod_{x \in \Omega} \exp\left(\beta \sum_{y \in \mathcal{N}_x} \delta\big(M(x) = M(y)\big)\right) \qquad (1)$$

$$L(x) \sim \frac{\exp\left(\rho D_{M(x)}^{L(x)}(x)\right)}{\sum_{l'=1}^{L} \exp\left(\rho D_{M(x)}^{l'}(x)\right)} \tag{2}$$

$$I^*(x) \sim \sum_{k=1}^{C_{L(x)}} \frac{w_{L(x),k}}{\sqrt{\left(2\pi\sigma_{L(x),k}^2\right)}} \exp\left[-\frac{(I^*(x) - \mu_{L(x),k})^2}{2\sigma_{L(x),k}^2}\right] \tag{3}$$

$$I(x) = I^*(x) \exp\left[-\sum_p c_p \psi_p(x)\right] \tag{4}$$

According to this model, segmentation can be cast as a Bayesian inference problem: given the image I and registered atlas segmentations $L_n$, the goal is to find the most likely segmentation $L$. Ideally, one would directly maximize $p(L|L_n, I)$ but this leads to an intractable integral over the model parameters ($\theta$ = means, variances, and bias field coefficients). Instead, we make the standard approximation that the posterior distribution of these parameters is heavily peaked around their mode $\hat{\theta}$. Then, we can first compute this mode ("point estimates") as $\hat{\theta} = \mathrm{argmax}_\theta p(\theta|I, L_n)$, and subsequently approximate $p(L|L_n, I) \sim p(L|L_n, I, \hat{\theta})$. To compute the point estimates, we resort to approximate inference, since the MRF leads to an intractable sum over the membership field $M$. We use a variational expectation maximization (VEM) algorithm, in which the posterior distribution of $M$ is approximated by a distribution that belongs to a restricted class of functions, specifically those that factorize over voxels (mean field approximation). Such approximation enables us to marginalize over $M$ when computing the point estimates. Further details can be found in [86].

Compared with [85, 86], a key difference is that some of our atlases do not provide sufficient contrast between gray and white matter to delineate the white matter surface. In those cases, we add an additional prior over $M$ that forces $M(x)$ not to be equal to the indices of such atlases over their white matter and cerebral cortex regions. In practice, this is easily implemented by making the (approximate) posterior $q_{x(M)}$ equal to zero for those atlases in those regions, in the E step of the VEM algorithm. This modification allows us to use a larger and not completely uniform training dataset, and to maximize the number of labels delineated in any given test images. Additionally, our segmentation algorithm does not assume that specific intensity distributions found in the training set are present in any new subject to be labeled. Instead it exploits the consistency of voxel intensities within target volume regions and the labels propagated. This is an important feature in the age range of 0-2 years, where myelination rapidly changes image contrast properties in a region-variant, disease-varying, and age-dependent manner.

The training dataset that we rely on for this task is a collection of 26 manually segmented T1-weighted images that are almost uniformly distributed in our age range of interest, with the exception of the newborn stage. Manual segmentation guidelines and 23 of the training examples were introduced in detail [42]. In addition, we recently augmented this set with another three examples. The segmentation algorithm allows for use of either the complete training dataset or a subset. Given an integer between 1 and the full training set size, we can automatically choose the members of that subset or *neighborhood*, using either the test subject's age or computing a mutual information-based image similarity between the test volume and the training subjects. We list the complete set of segmentation labels computed, along with corresponding FreeSurfer labels, in Table 1.

Given a multi-atlas segmentation approach, our segmentation framework requires that all atlas volumes be in the same spatial coordinate system as the test image. For this spatial normalization task, we rely on the Deformable Registration via Attribute Matching and Mutual-Saliency Weighting (DRAMMS) tool [88], which builds upon attribute matching and mutual-saliency weighting. We chose DRAMMS for its robust and accurate performance in the presence of image background noise, FOV differences, image appearance differences, and atlas-to-subject anatomical and age variations [89].

## 2.3 Surface Extraction

This step involves the tessellation of the gray matter-white matter boundary, automated topology correction [8, 90], and surface deformation following intensity gradients to optimally place the GM/WM and GM/CSF borders at the location where the greatest shift in intensity defines the transition between tissue class [5, 91, 92]. We found that, unlike in the case of adult image-processing solutions, the surface fitting performance gets more optimal when setting a relatively heavier weight on the volumetric image segmentations rather than intensity contrast information, due to less reliable contrast- and signal-to-noise ratios in infant acquisitions. Once the cortical models are complete, a number of deformable procedures are undertaken for surface inflation, registration to a spherical probabilistic atlas (based on individual cortical folding patterns to match cortical geometry across subjects), and creation of various surface-based data (ie, curvature, sulcal depth, and cortical thickness maps) [68, 91, 93]. Additionally, cortical parcellation information from an adult probabilistic atlas [41, 94] may also be extrapolated to the test subject at this stage.

## 2.4 Experiments

### 2.4.1 Datasets

#### 2.4.1.1 BCH_0-2 years

To quantify the accuracy of our results, we used a jackknifing (i.e. leave one out) strategy for images in our training dataset of 0-2 year-old infants, which were introduced and segmented accordingly [42] (Figure 3 and Appendix Table 3.) We retrospectively selected brain images of 26 infants, ranging from newborns to 2 year-old infants, scanned at Boston Children's Hospital (BCH) between 2009 and 2012. All MRI studies were clinically indicated. We screened clinical charts to ensure no genetic syndromes, no metabolic disorders, and no concerns of neurologic issues in qualifying subjects upon discharge. Additionally, it was mandatory that each subject's brain was deemed structurally normal by a pediatric neuroradiologist (PEG). As a common event in the post-delivery period, extracranial hematomas were not considered sufficient grounds for exclusion. The study was approved by the Committee on Clinical Investigation at BCH.

#### 2.4.1.2 BCHneo

For quantitative comparisons between our tool and others based solely on newborn datasets, we assembled a set of 17 healthy control neonates prospectively recruited at the BCH, independent of the training data set. Participating full-term neonates served for imaging purposes at 38.4±1.4 weeks of gestational age, solicited from the well-baby units at our collaborating hospitals and imaged at 28.9±10.5 days, with prior informed parental consent. They were all singletons with normal Apgar scores and no clinical concerns regarding perinatal brain injury or congenital or metabolic abnormalities. Of note, these data sets did not include corresponding full-brain manual segmentations.

### 2.4.2 Imaging acquisition

Scans of both BCH datasets were acquired using a 3 T MAGNETOM Trio Tim System or a 3 T MAGNETOM Skyra (Siemens Medical, Erlangen, Germany). Multi-echo volumetric magnetization prepared rapid gradient echo (MPRAGE) sequences [95] with volume navigators (vNav) for motion correction [96] (mocoMPRAGE) were obtained in the sagittal plane, at average image resolution of 1 mm$^3$, using a 32-channel adult head coil (see Appendix Table 3 for more acquisition details). *BCHneo* newborns also had T2-weighted image acquisitions completed in the same sessions. All subjects were imaged during natural sleep, and all images were assessed for quality. Those scans considered unsuitable for segmentation, due to degradation by motion or other artifacts, were not included in above-described cohorts.

*2.4.3 Skullstripping*

Despite the challenges of skullstripping in an infant population, our tool achieved >90% overlap with expert-delineated brain masks [79], as measured by the Dice overlap coefficient [97]. This performance is highly comparable to the long established adult skullstripping results of 94-96% [98]. For volumes A and B with manually and automatically outlined ROIs, respectively,

$$DICE(A,B) = \frac{2|A \cap B|}{|A|+|B|} \qquad (5)$$

where $[A \in S] = \{i \in I: A(i) \in S\}$. Representative skullstripping examples evaluated qualitatively are depicted in Figure 4, including original unprocessed *BCH_0-2yr* images and their intensity-normalized and skullstripped versions. For easier visualization, these were all aligned using affine registration to an unbiased spatial coordinate system [99-101].

*2.4.4 Automated vs manual volumetric segmentation*

We used *BCH_0-2yr*, which includes manual segmentations, to quantitatively characterize our segmentation accuracy. Each infant brain MRI was used as a test image and was segmented using a subset of the remaining atlases. We identified training datasets by postnatal age to investigate (similar to others [102] [43]) whether an age-related subset (vs the entire group) would be more accurate or efficient at the segmentation task. We varied the size of the training dataset from 1-25, proceeding from closest-in-age training subject to use of all the remaining datasets (excluding the test subject). Such age-dependent categories were motivated by a natural separation of the dataset, as well as by the fact that age matching of subjects would likely encourage more accurate segmentation results [102]. In summary, this resulted in running N = 26 × 25 = 650 segmentation experiments. For all label-to-label comparisons, we computed both Dice [97, 103] (corresponding to individual labels) and Generalized Dice (for overall accuracy) overlap coefficients of manually delineated and automatically outlined ROIs to quantify their agreement. For volumes A and B with manually and automatically outlined ROIs, respectively, the Generalized Dice score was computed as

$$DICE_{GEN}(A,B) = \frac{2|\cup_{s \in S}\{i \in I: A(i)=B(i)=s\}|}{|[A \in S]|+|[B \in S]|} \qquad (6)$$

where $[A \in S] = \{i \in I: A(i) \in S\}$. The Generalized Dice overlap coefficient was similarly defined in the generalized pair-wise multi-label Tanimoto Coefficient introduced by Crum *et al.* for fuzzy labels [104]. We compared such measurements across the entire dataset, in smaller age-related subsets, and also across all training dataset sizes. We only computed overlap coefficients if both manual and automated segmentation solutions existed for a given ROI.

*2.4.5 Automated volumetric segmentation comparison*

No other infant brain segmentation tool described in the literature is able to segment single-channel T1-weighted images in our proposed age range. Therefore, to compare our segmentation outcomes both qualitatively and quantitatively, we segmented our prospectively collected dataset of newborns (*BCHneo*), having both T1- (T1w) and T2-weighted (T2w) images for each subject, using two other publically available tools: iBEAT [46, 47, 51] and MANTIS [48]. We also processed T1-weighted structural images of the 40 subjects contained in the first release of The Developing Human Connectome Project [62].

*2.4.5.1 MANTIS*

MANTIS, the Morphologically Adaptive Neonatal Tissue Segmentation, extends the unified segmentation approach of tissue classification implemented in Statistical Parametric Mapping package (SPM [28]) to neonates. It utilizes a combination of unified segmentation, template adaptation via morphological segmentation tools and topological filtering, to segment the neonatal brain into eight tissue classes: cortical gray matter, white matter, deep nuclear gray matter, cerebellum, brainstem, cerebrospinal fluid (CSF), hippocampus and amygdala. This tool accepts brain-extracted T2-weighted images of newborns as inputs, so we used BET [70] processed T2w images from *BCHneo*. MANTIS does not make left/right hemispheric distinctions, so we combined our left and right labels (into a single label) for quantitative analysis. Label correspondences used for this analysis are outlined in Appendix Table 4.

*2.4.5.2 iBEAT*

iBEAT, the Infant Brain Extraction and Analysis Toolbox, integrates several major functions for infant image analysis, including image preprocessing, brain extraction, tissue segmentation, and brain labeling. For brain extraction, a learning-based meta-algorithm, integrating a group of brain extraction results generated by two existing brain extraction algorithms (BET [70] and BSE [81]) is used; for segmentation of infant brain tissues, a level-sets-based tissue segmentation algorithm utilizing multimodality information, a cortical thickness constraint, and a longitudinal consistency constraint is implemented; and for labeling regions of interest of infant brain images, the HAMMER (Hierarchical Attribute Matching Mechanism for Elastic Registration) [105] registration algorithm warps pre-labeled ROIs of a template to the infant brain image space. This tool accepts corresponding T1w and T2w images as inputs, producing both subcortical and cortical segmentation labels. Although finding the anatomic correspondence between the subcortical labels of our tool and iBEAT was straightforward, that was not the case for cortical labels. Therefore, we have only provided qualitative and quantitative comparisons of subcortical ROIs. Label correspondences used for this analysis are outlined in Appendix Tables 5.

*2.4.5.3 dHCP*

Even though this consortium processed T2w images of 40 newborn subjects in its first release, the corresponding T1w images were also made available. We processed these images in our new pipeline (using the five newborn atlases in our training data), comparing our segmentation results to those derived by the dHCP pipeline (incorporating BET [70] and drawEM using T2w MRIs [49]) relying on a set of mutually existing characterized labels. The original images of 0.8 x 0.8 x 0.8 mm$^3$ were downsampled to 1-mm isotropic resolution for our processing. Given that no ground-truth registration files are available that define spatial correspondence between the T1- and T2-weighted images and the set of overlapping segmentation labels common in the outputs of both pipelines is relatively low, we have only provided a qualitative segmentation comparison.

**Results:**

A detailed report of the inter-rater variability measures of the manual segmentation of our training set can be found elsewhere [42]. In brief, we reported results from two independent inter-rater variability studies. One showed that the worst performance (<60%), with respect to Dice overlap measures, was observed in the case of the L Amygdala. There were three structures for which the performance was 80-90% (L/R Putamen and R Thalamus), and one was >90% (L Thalamus). In the other study, L and R Accumbens performed worst (<50%), and there were nine labels where overlap was >80% (L/R Thalamus, L/R Caudate, L/R Putamen, Vermis, Midbrain, and Pons).These values represent an upper bound for the performance achievable by our automated pipeline.

*3.1 Qualitative segmentation evaluation*

We first demonstrated the quality of our automated segmentation using a set of five representative and variably aged subjects (newborn, 8 mo, 12 mo, 16 mo, and 18 mo), displaying both manually and automatically labeled brain images. Figure 5 shows selected snapshots of these subjects in coronal views, with the input image, the manually segmented solution, and its outline (overlain on input image), plus the corresponding segmentation and its outline (overlain on input image) in 5 respective columns. In the first row, input images of a newborn clearly display the reverse intensity contrast of an adult. However, in other subjects of the remaining rows, the intensity contrast more closely resembled that of an adult, albeit with unmyelinated areas still visible. In all of these images, GM/WM boundaries and subcortical region segmentations demonstrate high levels of correspondence. Of note, WM of cerebellum was not included in manual segmentation of the newborn or the 12 month-old and thus is missing from the second-row display.

In Figure 6, we have shown another set of representative images from five other subjects aged 2, 3, 5, 6, and 9 months. In these subjects, manual segmentations did not include the GM/WM boundaries due to uncertain contrast intensity. However, the automated tool still recovered white matter segmentation labels with acceptable accuracy in the majority of the cases, when comparing images with WM-labeled training subjects of similar ages.

*3.2 Quantitative segmentation evaluation*

Figure 7 is a graph of Generalized Dice overlap coefficients for all 26 subjects evaluated in *BCH_0-2yr* over training set sizes of 1-25, selected by age. The subjects are presented by age in ascending order (dark blue indicating the youngest subject and bright yellow indicating the oldest). The highest measurements are attributable to subjects in the middle of our age range of interest, whereas the lowest are those of newborns. We must also acknowledge that in all cases, use of a subset rather than the complete training set, yielded better overall segmentation performance. This may be explained by age differences in the training dataset. Such trends are even more obvious when we display our overlap measures for five non-overlapping age groups across the increasing training set sizes: newborns (N=5), 2-4 mo (N=4), 5-8 mo (N=5), 9-14 mo (N=6), and 15-18 mo (N=6). Figure 8 shows the average and Figure 9 the maximum Generalized Dice measures for these age groups. In the former, the highest values reached 0.83; but in the latter, a value of nearly 0.94 was recorded.

To quantify which neighborhood size in the label fusion algorithm yields the best segmentation performance, we have shown (Figure 10) the number of times a particular neighborhood size prevailed as best performer (using overall Dice overlap coefficients) across the five age groups. There is ample clustering of winning numbers under the size 5 training dataset. Figure 11 displays the Generalized Dice overlap score versus age-at-scan computed on the training data set for this neighborhood size.

In Supplementary Figures 1-3, average Dice scores per segmentation ROIs are depicted over all subjects in *BCH_0-2yr*, with respect to training set sizes of 1-12 (selected by age). The highest score was consistently achieved in L/R Thalamus and Pons regions.

*3.3 Qualitative and quantitative comparisons with other tools*

As discussed above, a fair direct comparison of these tools/datasets (with differing input image requirements and differing segmentation label sets) is not feasible. Nevertheless, we examined their segmentation outcomes to better appreciate the inherent disparities and the difficulty in absolute ranking of tools. Note, no manual labels were available for the below testing data sets, so the segmentation outcomes were just directly compared to each other and not to ground truth.

We ran MANTIS [48] on T2w images of 17 newborns (GA, 38.7±1.2 wks; age at scan, 28.2±11.4 days) and our segmentation pipeline using corresponding T1w images, comparing commonly identified segmentation labels. The list of such ROIs included cerebral cortex, cerebral white matter, deep grey matter, hippocampus, amygdala, cerebellum, and brainstem. MANTIS does not distinguish between left and right hemispheric labels, so for sake of comparison, we merged our hemispheric labels. We have shown coronal views of input images and segmentation outcomes (affinely aligned for visualization purposes) in Figure 12. In Figure 13, Dice overlap measures computed for corresponding labels are plotted for each subject. The overall match between these segmentation solutions was 0.7-0.8, with brainstem, cerebellum, and deep gray matter closer to 0.8 and hippocampus and amygdala often <0.5.

We also ran iBEAT [51] on a set of twelve matching T1w and T2w images of newborns (GA, 38.7±1.4 wks; age at scan, 29.75±11 days) and our segmentation pipeline on their corresponding T1w images and compared the commonly identified labels in the outcomes. In our qualitative and quantitative comparison we only include the commonly defined subcortical labels: left/right thalamus, caudate, putamen, pallidum, hippocampus, and amygdala. We display a coronal view of the input T1w images and the segmentation outcomes (affinely aligned for visualization purposes) in Figure 14. Figure 15 shows Dice overlap measures computed between corresponding labels for each subject. There are no labels that seem to consistently perform better or worse, but the amygdala, putamen and hippocampus produce higher matches in some cases. The overall match between these labels, however, is lower, around 0.6.

We used our new segmentation tool to process images of the 40 subjects in the 1$^{st}$ dHCP data release [62]. Coronal views of the input T1w images, our skull-stripping solutions, and our segmentation outcomes (affinely aligned for visualization purposes) are shown in Figures 16 and 17, with all commonly identified ROIs: left/right cortical gray matter, cortical white matter, ventricles, hippocampus, amygdala, caudate, thalamus, and lateral ventricles, as well as the cerebellum and brainstem. Figure 18 shows T2w input images of the dHCP and their segmentation solutions. Overall, the dHCP cortical segmentations seem slightly more accurate. One key explanation for this is the discrepancy in qualities of T1w and T2w input images (for example, second row 1$^{st}$ or third row 3$^{rd}$ and 4$^{th}$ images at tops of Figures 17 and 18).

Figure 19 displays Mean Dice overlap measures computed on this dataset per segmentation labels between our and the dHCP processing pipeline. The nomenclature originates from the data released by the dHCP consortium: "all" identifying 13 labels in total and "tissue" segmentation labels referring to more detailed segmentation results, but only 7 of them overlapping with our definitions.

*3.4 Surface extraction*

Due to lack of ground-truth surface reconstruction of our input images we first display surfaces for qualitative evaluation. In Figure 20, representative surface models of five subjects (newborn, 8, 12, 16, and 18 months old) are shown. The surfaces are those of white matter, pia, and a spherical representation with curvature-map overlay. In addition to a qualitative evaluation, we also performed a set of quantitative quality control tests. On a scale of 0-5, two independent experts scored the quality of white matter and pial surface reconstructions of the same five subjects appearing in Figure 20, deducting scores from maximum for errors such as holes, mislabeled ventricles, dura grabbing, or missed gyri. Both evaluations resulted in average surface scores of 3, with standard deviations of 0.7 and 0.8. Detailed outcomes of this analysis are provided in Supplementary Figure 4.

On the *dHCP* data set we also compared the surfaces provided by the *dHCP* pipeline to our proposed outcomes. Figure 21 displays a comparison between surface measurements per hemispheres. We computed the mean absolute distance between the two sets of solutions as well as the mean sulcal depth, cortical thickness and curvatures differences. In summary, these differences were on average 1.17mm, -0.5, 0.9 and -0.09, respectively. The more detailed, per cortical parcellation label comparison of these measurements is included in Supplementary Figure 5.

Additionally, we randomly selected 12 and 10 subjects, respectively, from the *BCH_0-2 years* and *dHCP* data sets and a trained expert drew points on the white matter and pial surfaces based on their T1 weighted MRI. On average 75 control points were placed on both of these surfaces, on both hemispheres (see Figure 22 for an example). We then computed the shortest distance between these points and our computed surfaces. Additionally, in the case of the *dHCP* dataset, we also computed distances between the manually drawn points and surfaces provided by the *dHCP* minimal processing pipeline. The mean and standard deviation of the absolute value of these measurements are included in Table 2. The white matter surfaces tend to be slightly higher in the case of our proposed method, while in the case of the *dHCP* pipeline it is the pial surface differences that are higher. On the studied *dHCP* data set, the minimal processing pipeline outperforms our solutions with a statistically significant difference at the 5% significance level.

**Discussion:**

*4.1 General comments*

We have introduced an automated segmentation and surface extraction pipeline for image processing in infants designed to accommodate clinically acquired infant brain MRI data from a population of 0-2 year-olds. To our knowledge, there is no algorithm or computational pipeline capable of consistently handling single time-point T1-weighted MR images of subjects within this postnatal period, producing full-brain volumetric segmentations and surface extractions. The innovative aspect of this method resides in the adaptation of three key methodological solutions that aggregate infant MRIs in a pipeline for thorough evaluation. These key components are a manually segmented training dataset for the age range of interest, a robust skull-stripping algorithm, and a multi-atlas label-fusion segmentation framework benefitting from information encoded in the training set and the surface extractions. The primary advantage of our algorithm is that it can be optimized for any age. Indeed, the pipeline selects a subset of the training dataset most similar to the target subject to generate segmentations and surfaces.

Through our experimentation, we presented both qualitative and quantitative evaluations to characterize the performance of our tool, showing that its overall functioning was consistent across the target age range; and its accuracy (as measured by Dice and Generalized Dice coefficients) was high for such a difficult task. The highest mean and maximum generalized Dice overlap coefficient scores were obtained for the 2-8 mo subset and the lowest for the newborns. We also investigated whether age-dependent sub-grouping of the manually segmented datasets would be beneficial and found that a training dataset of few atlases close in age to test subjects was optimal. Observations on diminishing returns and worsening performance of an enlarging training dataset may have validity in this instance, serving to increase the diversity of the training data. The more training subjects used, the larger the age disparity between the test subject and at least some training subjects. However, those subjects deviating substantially from the test subject, may not contribute significantly to segmentation accuracy owing to information conflicts with closer-in-age subjects and possibly higher registration inaccuracies. Note also that we ran the same number of experiments for training-set size selection, using mutual information (MI)-based criteria and focusing on image similarity, as opposed to aligning subjects with atlases by age. In a majority of cases, age-related selection performed superiorly, which is why we omitted performance metrics for the MI-based experiments conducted. We believe that performance discrepancies are readily explained by the fact that for our normal control group, age was a robust and reliable parameter determining structural similarities of input images. However, in the presence of disease this may not be true, so image-based selection criteria will also be available in our forthcoming software release.

We likewise compared our segmentation solutions both qualitatively and quantitatively to three other publically available algorithms based on newborn training datasets. The overall correspondence between such outcomes was generally good but was also quite varied, showing perhaps the best correspondence with dHCP results. However, due to differences in label definitions, as well as shifting requirements and quality of input images, such comparisons should be further investigated moving forward.

Motion artifacts have a great impact on infant image quality. It is almost impossible, however, to quantify the amount of motion in a scan if no steps had been taken during the acquisition stage to save related information (such as head tracking). Given that we used retrospectively selected data sets in this study, we had no such information at hand. As an alternative, we computed a reference-free measure of image sharpness, the Tenengrad metric from [106], in order to quantitatively characterize the quality of our data sets. We computed this metric in the common affine (visualization) space based on the middle slice of the image volume. First we point out the close relationship between the age-at-scan and the Tenengrad metric displayed in Figure 23. As the former gets higher, the sharpness metric also tends to increase. Second, Figure 24 displays the Generalized Dice score for each of training subjects using 4 and 5 as training neighborhood sizes vs the input image volumes' Tenengrad metric (fmeasure). This figure demonstrates a clear tendency for a higher Generalized Dice score associated with a higher sharpness metric (and higher age-at-scan).

The Tenengrad image sharpness metric for the dHCP newborns on the T1-weighted images was in the same range as that of those of the newborn training subjects. Figure 25 displays this score along with the Generalized Dice score using a training neighborhood size of 5: "all" segmentation labels (in red) and common-with-our-pipeline "tissue" labels (in blue).

The segmentation and image processing pipeline described in this paper will be distributed in source and binary format under the existing FreeSurfer platform and under a modified MIT-style license [107], in conjunction with our training dataset.

*4.2 Limitations*

Current limitations of our image processing solution stem from the fact that our proposed pipeline as yet does not accommodate T2w MRI images, which typically confer higher CNRs for infants up to 6 months of age [108]. This limits our ability to directly compare the performance of our application with that of existing tools, particularly in terms of the newborn sub-population. At the same time, we also fill a void in the literature, providing a tool for researchers and clinicians who use more clinically practical 2D T2w images (rather than longer, high-resolution isotropic T2w images) in regular patients but generally acquire faster volumetric T1w images for clinical studies. The current pipeline resamples all input images to be 1mm isotropic in order to match the resolution of the training data sets. In the future, we aim to remove this constraint by obtaining higher resolution manually segmented data sets for both training and validation. In their present state, our tools may already generate a set of cortical parcellations analogous to those of the FreeSurfer adult pipeline, but we did not characterize them in this paper. Finally, the current training dataset is missing some GM/WM boundary descriptions. We feel that increasing the number of training subjects, gathering full GM/WM segmentations across the entire age range will further increase the accuracy and consistency of our segmentation and surface extraction outcomes.

*4.3 Future initiatives*

The current pipeline provides an excellent platform for future extensions as follows: (i) A planned extension of our segmentation and skullstripping training datasets using T2w samples, potentially enhancing segmentation accuracy in newborn populations and allowing direct comparisons of our performance with outcomes of other publically available image-processing tools and multi-site datasets: The emergence and public sharing of datasets through initiatives such as dHCP (available at http://www.developingconnectome.org/) and MICCAI challenge datasets (available at http://neobrains12.isi.uu.nl, http://iseg2017.web.unc.edu/ and http://iseg2019.web.unc.edu/) will be instrumental in this regard; (ii) Full use of the thoroughly tested FreeSurfer framework (consistently handling analysis of longitudinal images [100, 109]) to potentially increase the sensitivity/specificity of follow-up group analyses and require fewer subjects to detect comparable effect sizes [100, 109]: The FreeSurfer implementation of longitudinal processing is somewhat similar to that of Shi *et al.* [110], without favoring any of the time points, which may lead to bias and encourage spurious effects [109, 111]; (iii) Annotation of cortical parcellation areas, specifically in our population of interest, on surfaces that are currently extracted from our volumes and include those references within our pipeline: This would provide a new infant-specific set of labels comparable to the recent release by Alexander *et al.* defined on T2w images for newborns [3, 112]; (iv) Emphasis on performance in cortical surface placement: Cortical thickness is a powerful biomarker that has been used in many clinical studies to assess of a variety of neurologic and neurodevelopmental outcomes [113-117]. In the future, we plan to manually estimate cortical thickness, comparing resultant values with measures reported in the literature; and (v) The emergence of neural network-based image analysis frameworks, in particular solutions estimating deformable templates, creates an exciting opportunity to incorporate efficient learning solutions, such as conditional atlases [118], into our infant brain analysis pipeline.

**Acknowledgements:**

Support for this research was provided in part by NIH/NICHD grants 1K99HD061485-01A1, R00 HD061485-03 (LZ); A. A. Martinos Center Computing facilities: NIH S10RR023401, S10RR019307, S10RR019254, S10RR023043; NIH/NIBIB grants P41EB015896, 1R01EB023281, R01EB006758, R21EB018907, R01EB019956, NIH/NIA grants 1R56AG064027, 5R01AG008122, R01AG016495, NIH/NIDDKD grant 1-R21-DK-108277-01, NIH/NINDS R01NS0525851, R21NS072652, R01NS070963, R01NS083534, 5U01NS086625, the NIH Blueprint for Neuroscience Research (5U01-MH093765), part of the multi-institutional Human Connectome Project (BF); the European Research Council (Starting Grant 677697, project BUNGEE-TOOLS) (JEI); NIH R01HD076258 (PEG); NIH R01EB014947 (YO, PEG); Thrasher Early Career Development Grant (YO); as well as partially from Abbott Nutrition through the Center for Nutrition, Learning, and Memory at the University of Illinois (trial being registered at clinicaltrials.gov as NCT02058225). In addition, BF has a financial interest in CorticoMetrics, a company whose medical pursuits focus on brain imaging and measurement technologies. BF's interests were reviewed and are managed by Massachusetts General Hospital and Partners HealthCare in accordance with their conflict-of-interest policies. Finally, the authors would like to acknowledge the meticulous and detail-oriented work of several students and research assistants who assisted us with analysis of the above described dataset: Rutvi Vias, Christopher Ha, Lucy Schlink, and Ngo Giang-Chau.

*Compatible Version of M-CRIB Neonatal Parcellated Whole Brain Atlas: The M-CRIB 2.0.* . Front Neurosci., 2019. **13**(34).
113. McCauley, S.R., et al., *Patterns of Cortical Thinning in Relation to Event-Based Prospective Memory Performance Three Months after Moderate to Severe Traumatic Brain Injury in Children.* Developmental neuropsychology, 2010. **35**(3): p. 318-332.
114. Almeida, L.G., et al., *Reduced right frontal cortical thickness in children, adolescents and adults with ADHD and its correlation to clinical variables: a cross-sectional study.* Journal of Psychiatric Research, 2010. **44**: p. 1214-1223.
115. Kirk, G.R., et al., *Regionally Specific Cortical Thinning in Children with Sickle Cell Disease.* Cerebral Cortex, 2009. **19**(7): p. 1549--1556.
116. Wolosin, S.M., et al., *Abnormal Cerebral Cortex Structure in Children with ADHD.* Human Brain Mapping, 2009. **30**(1).
117. Merkley, T.L., et al., *SHORT COMMUNICATION: Diffuse Changes in Cortical Thickness in Pediatric Moderate-to-Severe Traumatic Brain Injury.* Journal of Neurotrauma, 2008. **25**(11).
118. Dalca, A.V., et al., *Learning Conditional Deformable Templates with Convolutional Networks*. 2019: arXiv.org > cs > arXiv:1908.02738.

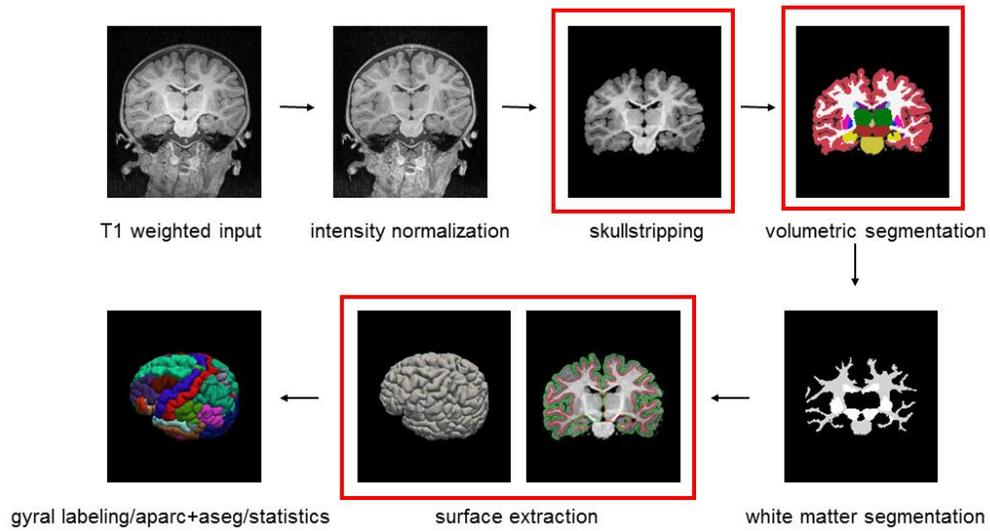

Figure 1: Major image processing steps in the standard FreeSurfer reconall pipeline. Red boxes indicate the ones that are different and were specifically modified in the case of the infant-specific tools.

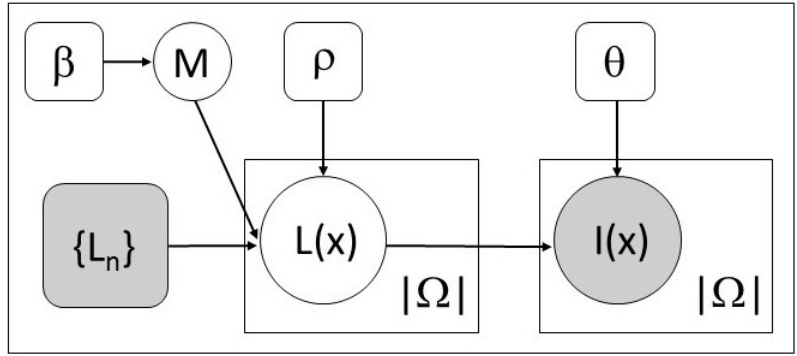

Figure 2: The proposed graphical model for our multi-atlas label fusion tool. Plates indicate replication, shaded variables are observed.

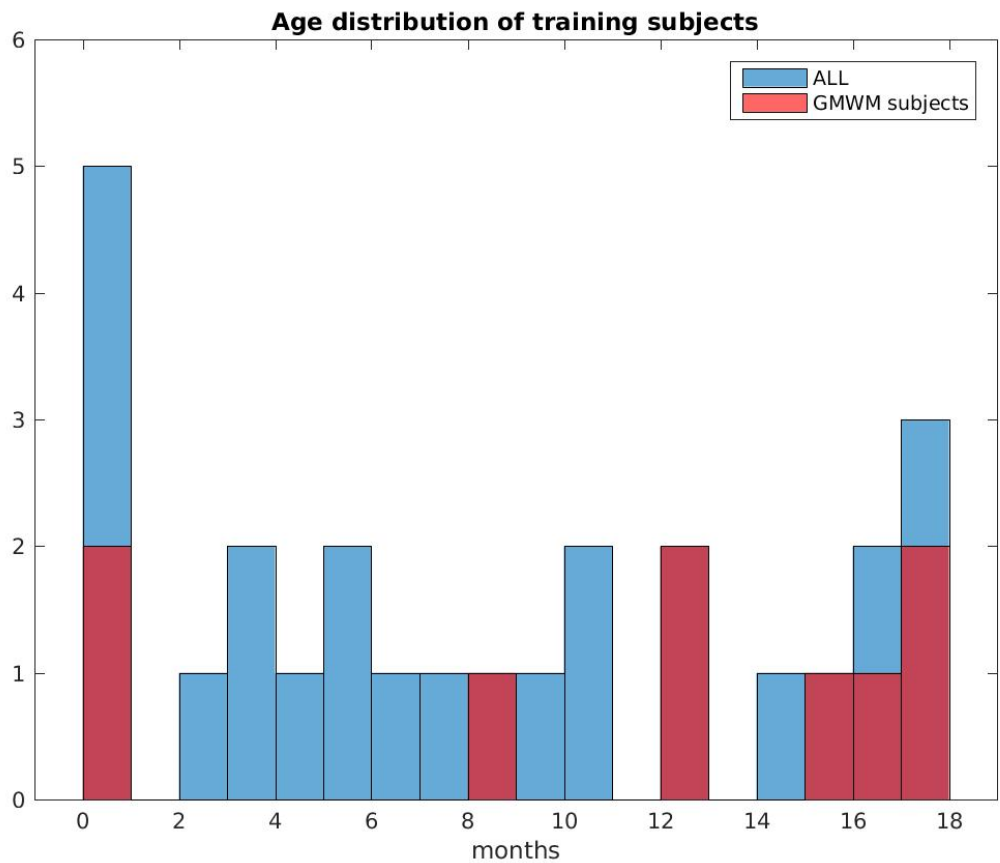

Figure 3: Age distribution at scan of the twenty-six subjects in the training data set. Red color indicates data samples that had GM/WM separation drawn by the manual labelers.

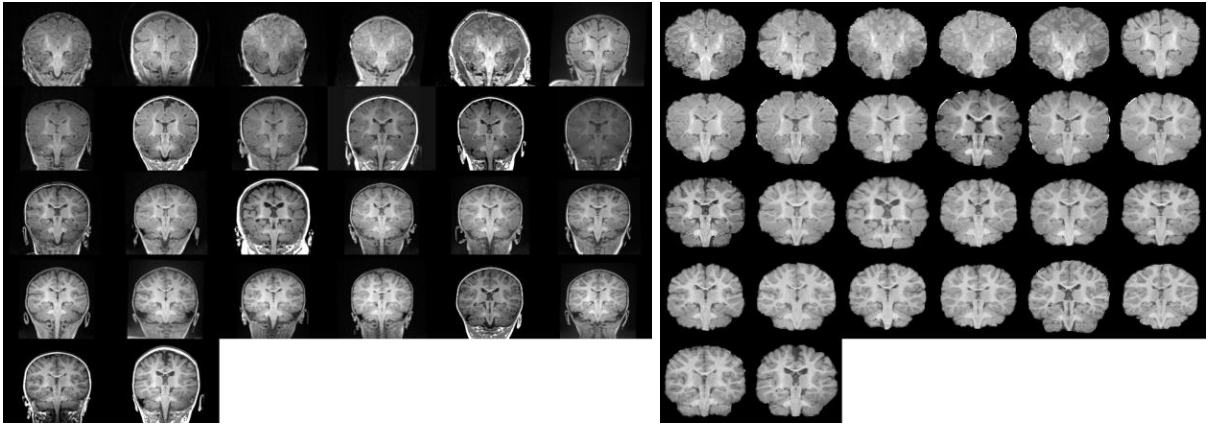

Figure 4: Skull-stripping results: (left) unprocessed images from *BCH_0-2yr* data set and (right) intensity normalized and skull-stripped results. Both sets are age sorted, displayed in coronal view and aligned using affine registration to an unbiased spatial coordinate space for easier visualization.

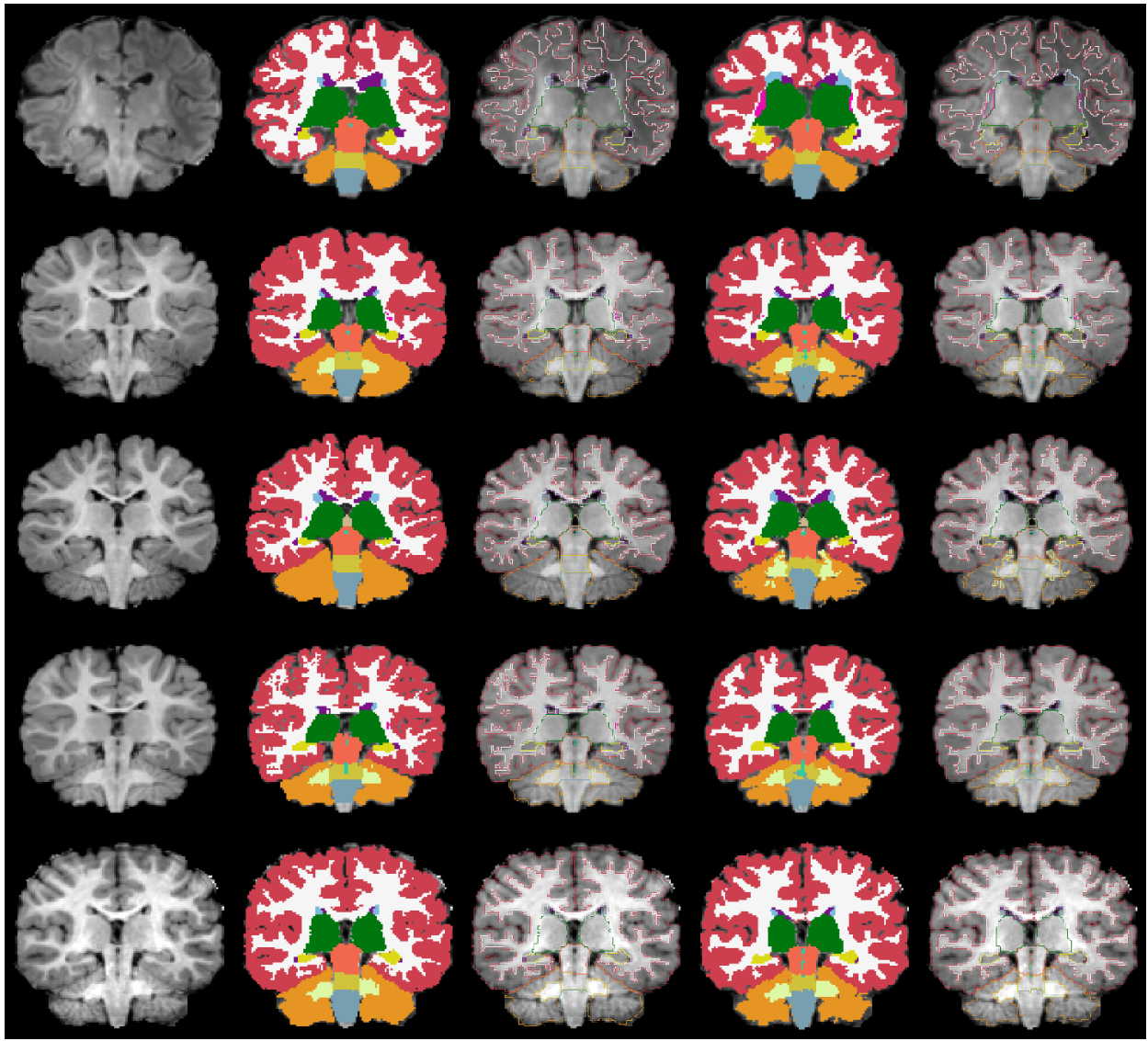

Figure 5: Five automated segmentation examples where manual segmentation also contained GM/WM separation: (from top to bottom) newborn, 8mo, 12mo, 16mo, 18mo. From left to right: normalized and skullstripped T1-weighted input image, manual segmentation, manual segmentation outline, automated segmentation, automated segmentation outline. All segmentations (or their outlines) are overlaid on the normalized and skullstripped T1-weighted input image. The segmentation colors correspond to the default Freesurfer colortable.

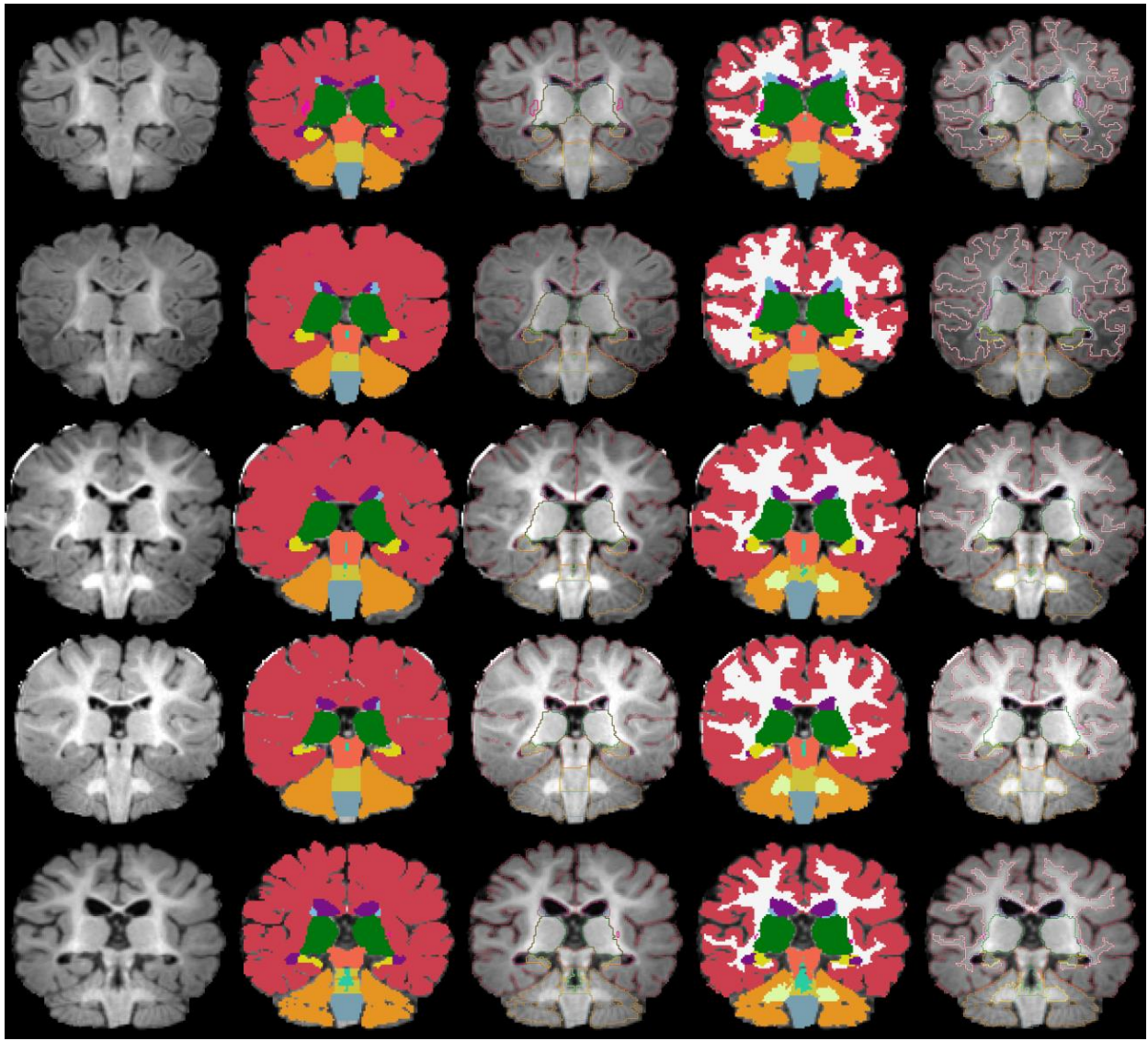

Figure 6: Five automated segmentation examples where manual segmentation did not contain GM/WM separation: (from top to bottom) 2mo, 3mo, 5mo, 6mo, 9mo. From left to right: normalized and skull-stripped T1-weighted input image, manual segmentation, manual segmentation outline, automated segmentation, automated segmentation outline. All segmentations (or their outlines) are overlaid on the normalized and skull-stripped T1-weighted input image. The segmentation colors correspond to the default Freesurfer colortable.

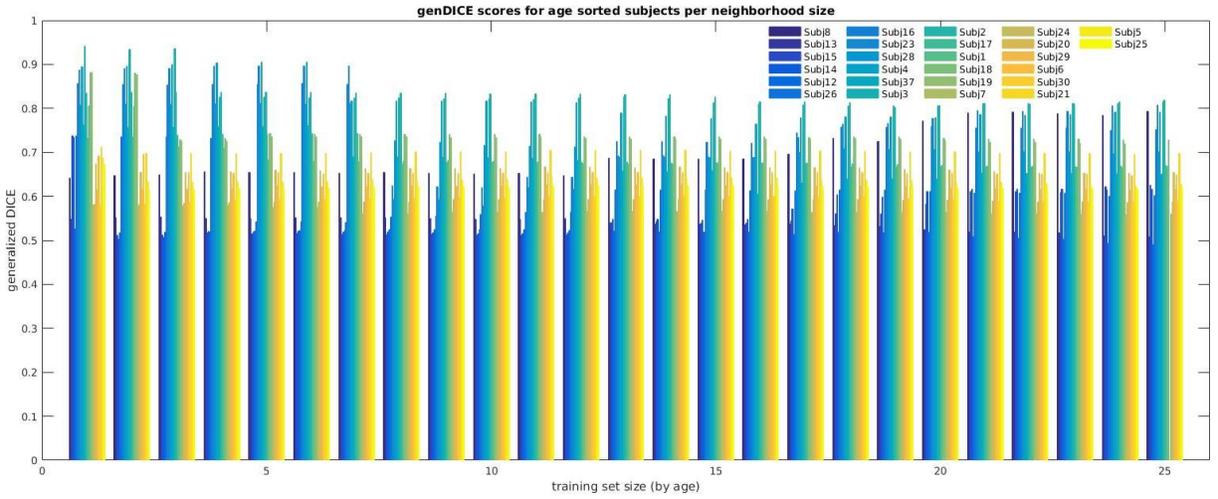

Figure 7: Generalized Dice overlap coefficient summary for all subjects and training set sizes (selected by age). The generalized Dice coefficients are displayed for training set sizes 1-25 for all of our subjects, in an age-sorted manner: Subj8 (newborn) → Subj 25 (18 mo).

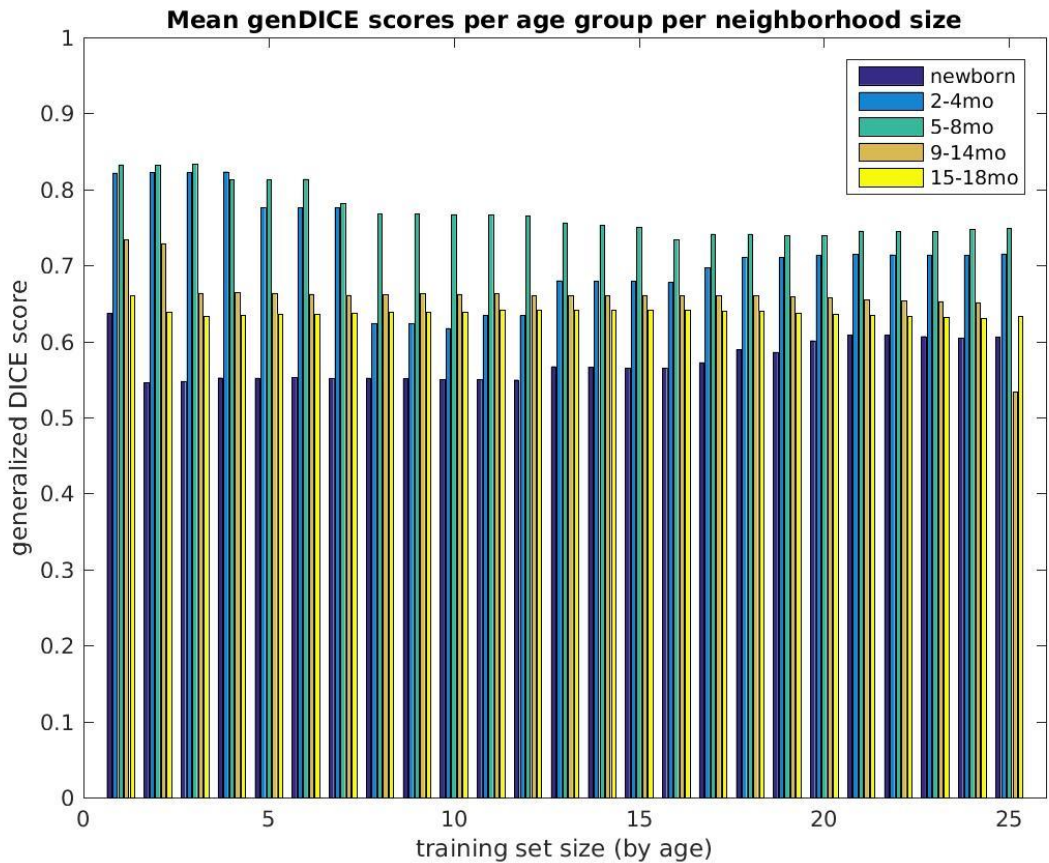

Figure 8: Mean generalized Dice overlap coefficient summary over all training set sizes (selected by age) for all subjects, grouped into five non-overlapping age groups (newborns (N=5), 2-4 month (N=4), 5-8 month (N=5), 9-14 month (N=6) and 15-18 month olds (N=6)). The measures are displayed for training set sizes 1-25.

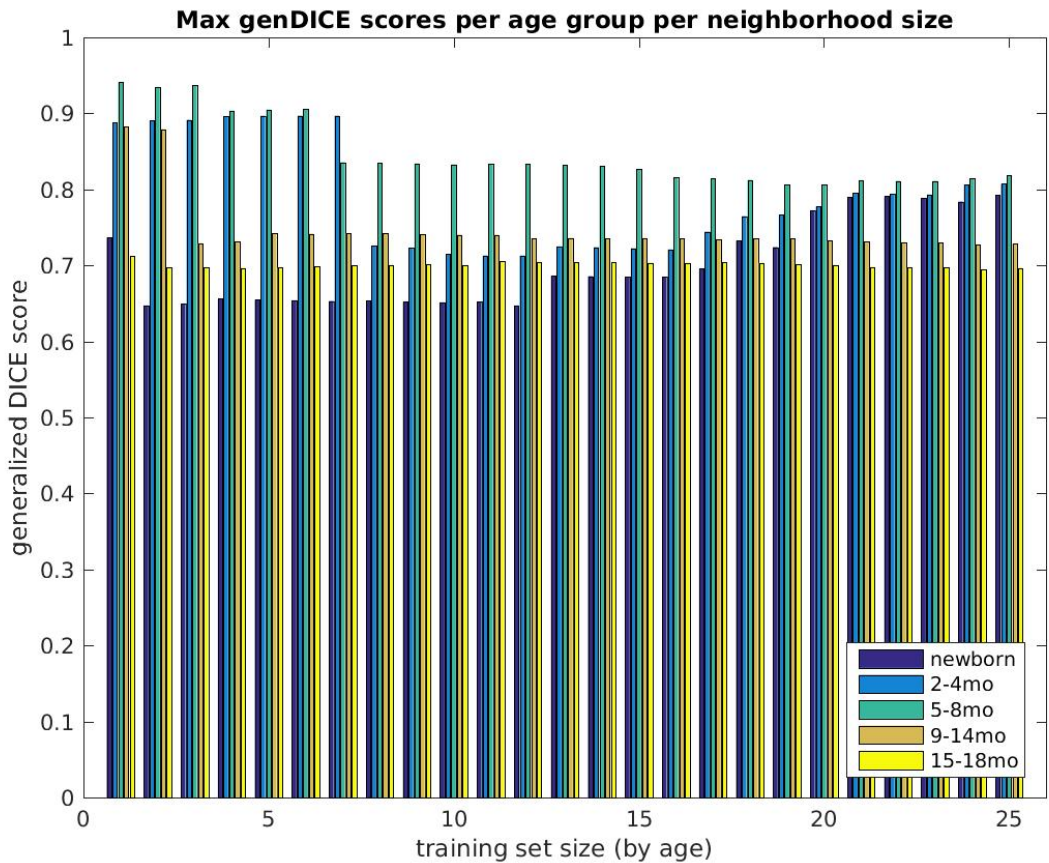

Figure 9: Maximum generalized Dice overlap coefficient over all training set sizes (selected by age) for all subjects grouped into five non-overlapping age groups (newborns (N=5), 2-4 month (N=4), 5-8 month (N=5), 9-14 month (N=6) and 15-18 month olds (N=6)). The measures are displayed for training set sizes 1-25.

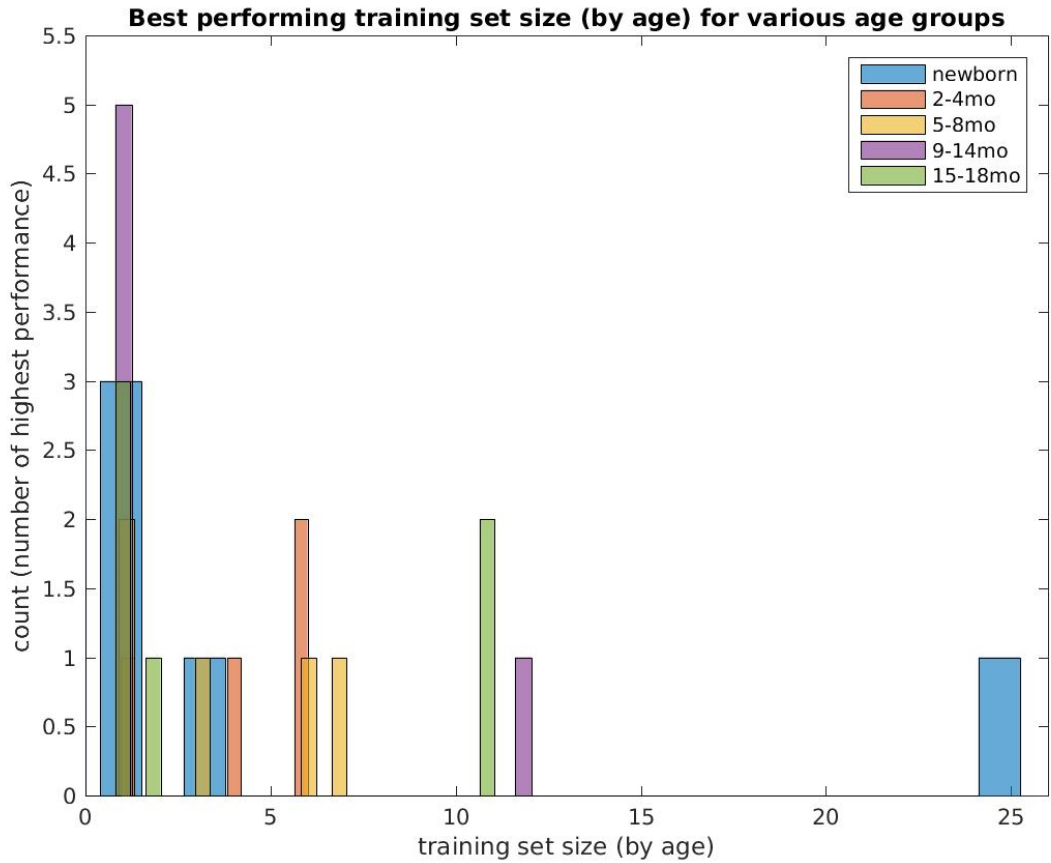

Figure 10: Best performing training set sizes (selected by age) computed using generalized Dice coefficients in five non-overlapping age categories (newborns (N=5), 2-4 month (N=4), 5-8 month (N=5), 9-14 month (N=6) and 15-18 month olds (N=6)).

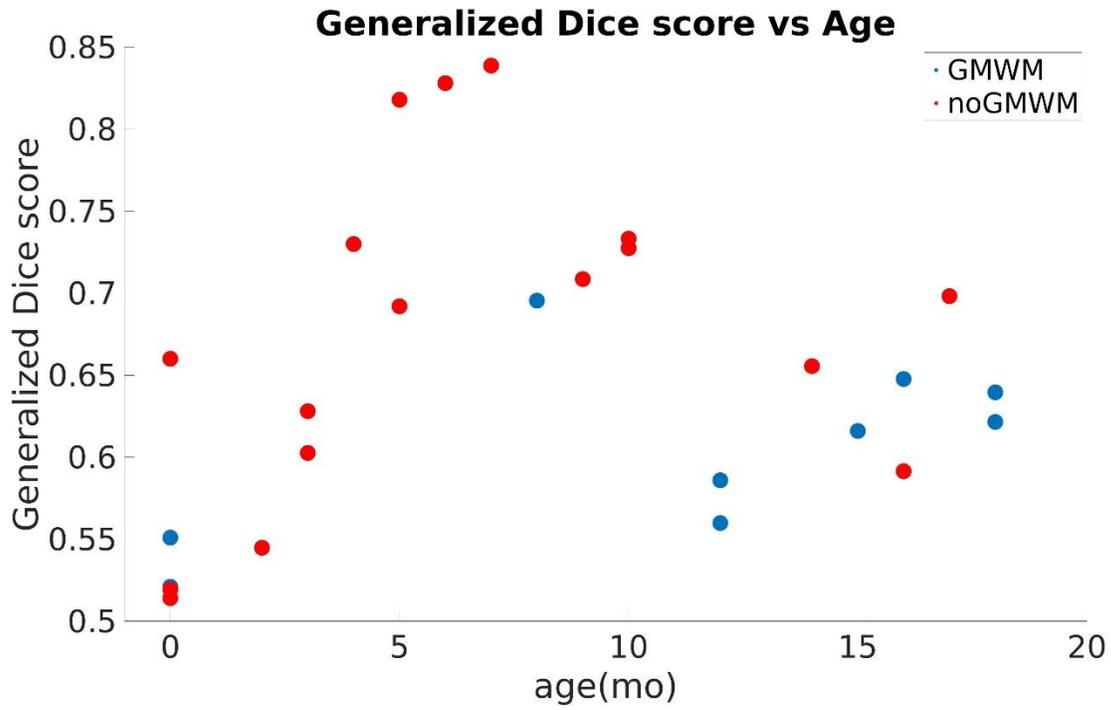

Figure 11: Generalized Dice score vs age-at-scan computed on the training data set for neighborhood size 5.

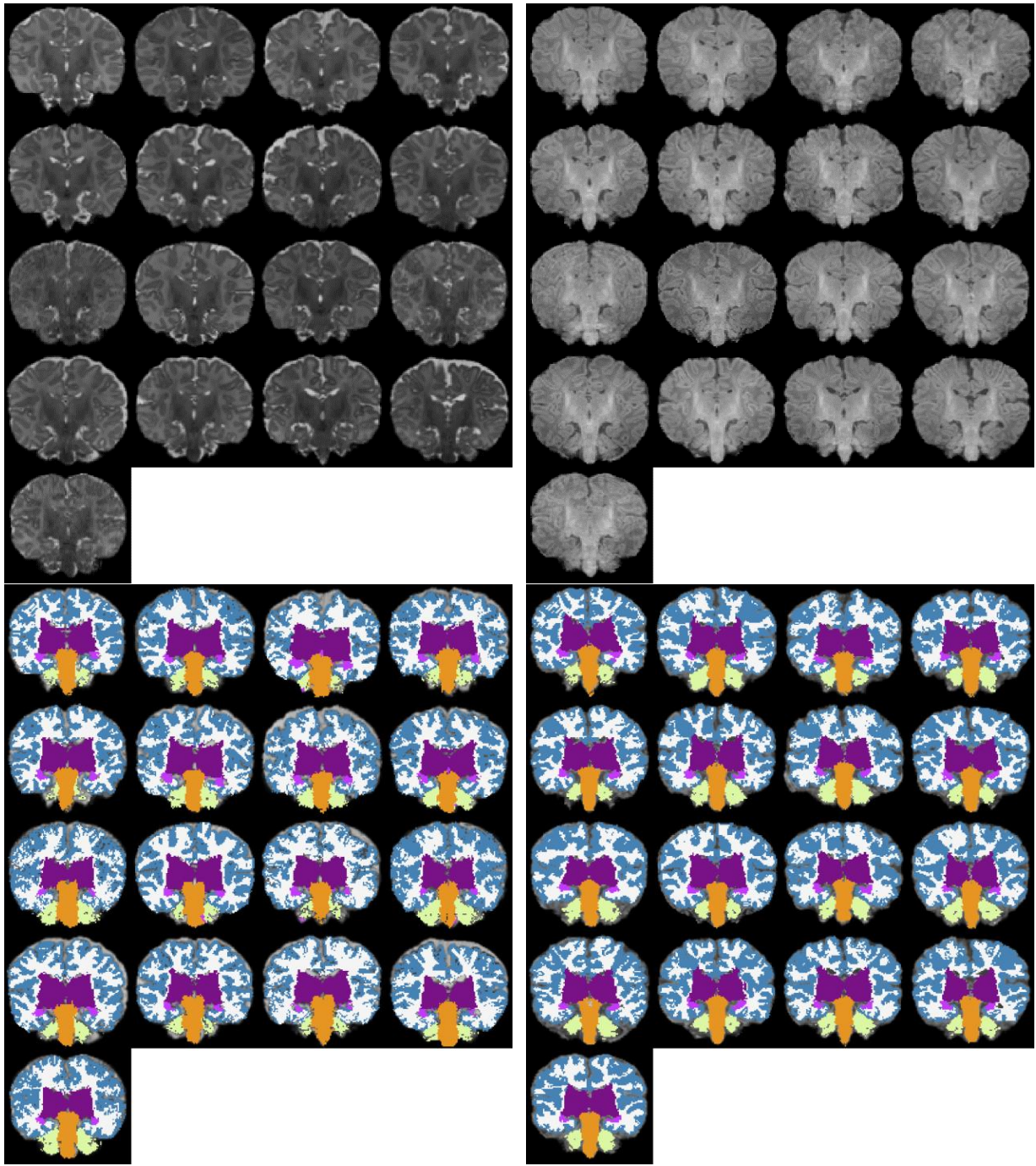

Figure 12: MANTIS segmentation comparison: (top left) T2w input images (bottom left) MANTIS segmentations, (top right) corresponding T1w input images (bottom right) our segmentation outcome after grouping left / right hemisphere labels together. The list of commonly identified labels are: cerebral cortex, cerebral white matter, deep grey matter, hippocampus, amygdala, cerebellum and brainstem. For more detailed label correspondences see Appendix Table 4.

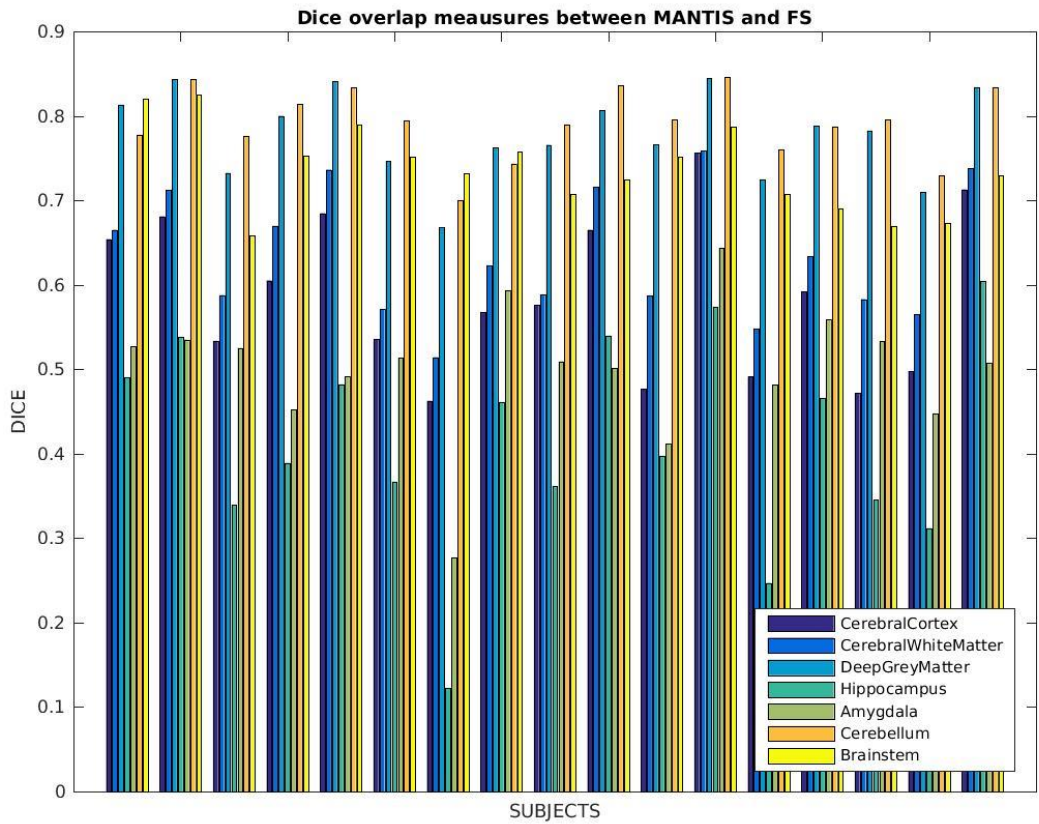

Figure 13: Dice coefficients computed between our segmentations and MANTIS for labels that are commonly identified by these tools: cerebral cortex, cerebral white matter, deep grey matter, hippocampus, amygdala, cerebellum and brainstem.

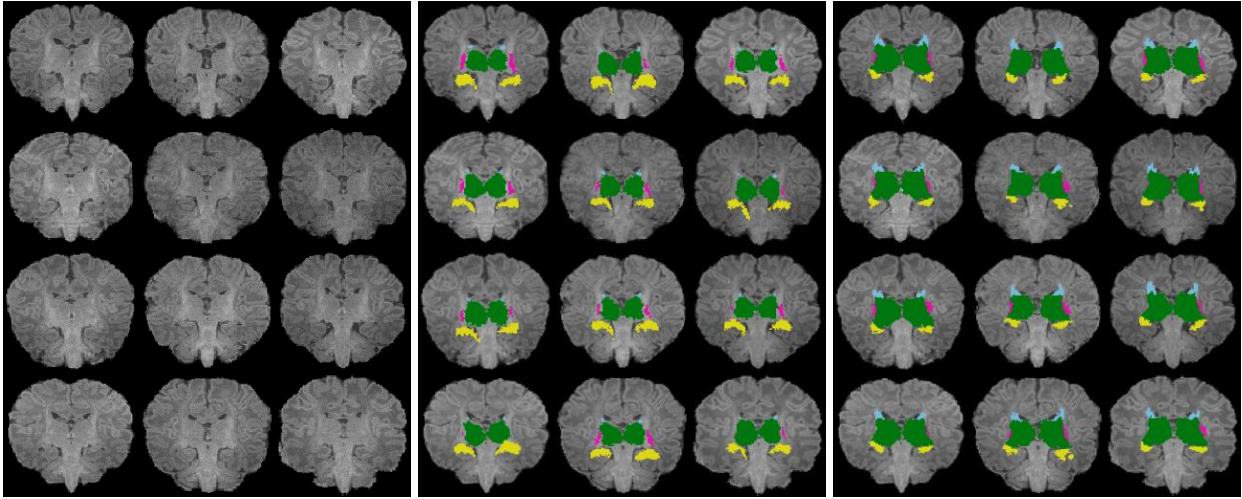

Figure 14: iBEAT subcortical segmentation comparison: (left) T1w input images (middle) iBEAT segmentations, (right) our segmentation outcome. The list of commonly identified labels are: left / right thalamus, caudate, putamen, pallidum, hippocampus and amygdala. For more detailed label correspondences see Appendix Table 5.

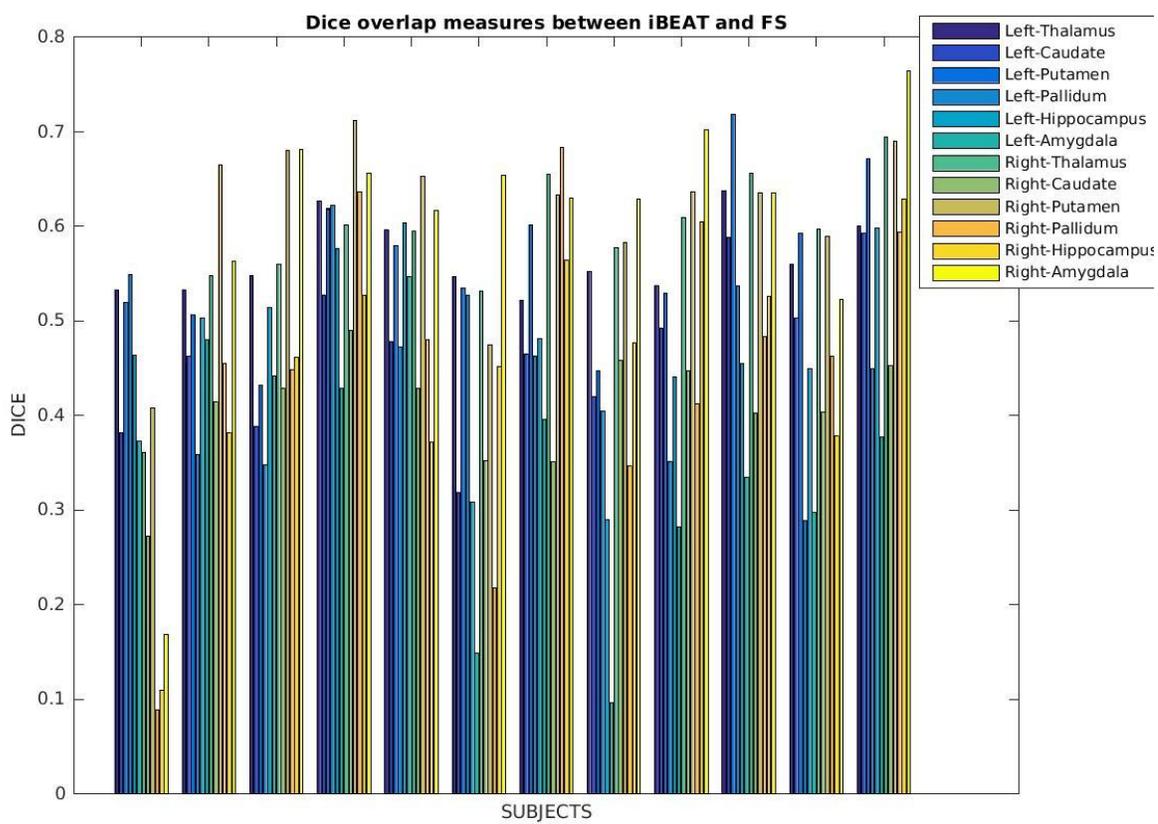

Figure 15: Dice coefficients computed between our segmentations and iBEAT for labels that are commonly identified by these tools: left / right thalamus, caudate, putamen, pallidum, hippocampus and amygdala.

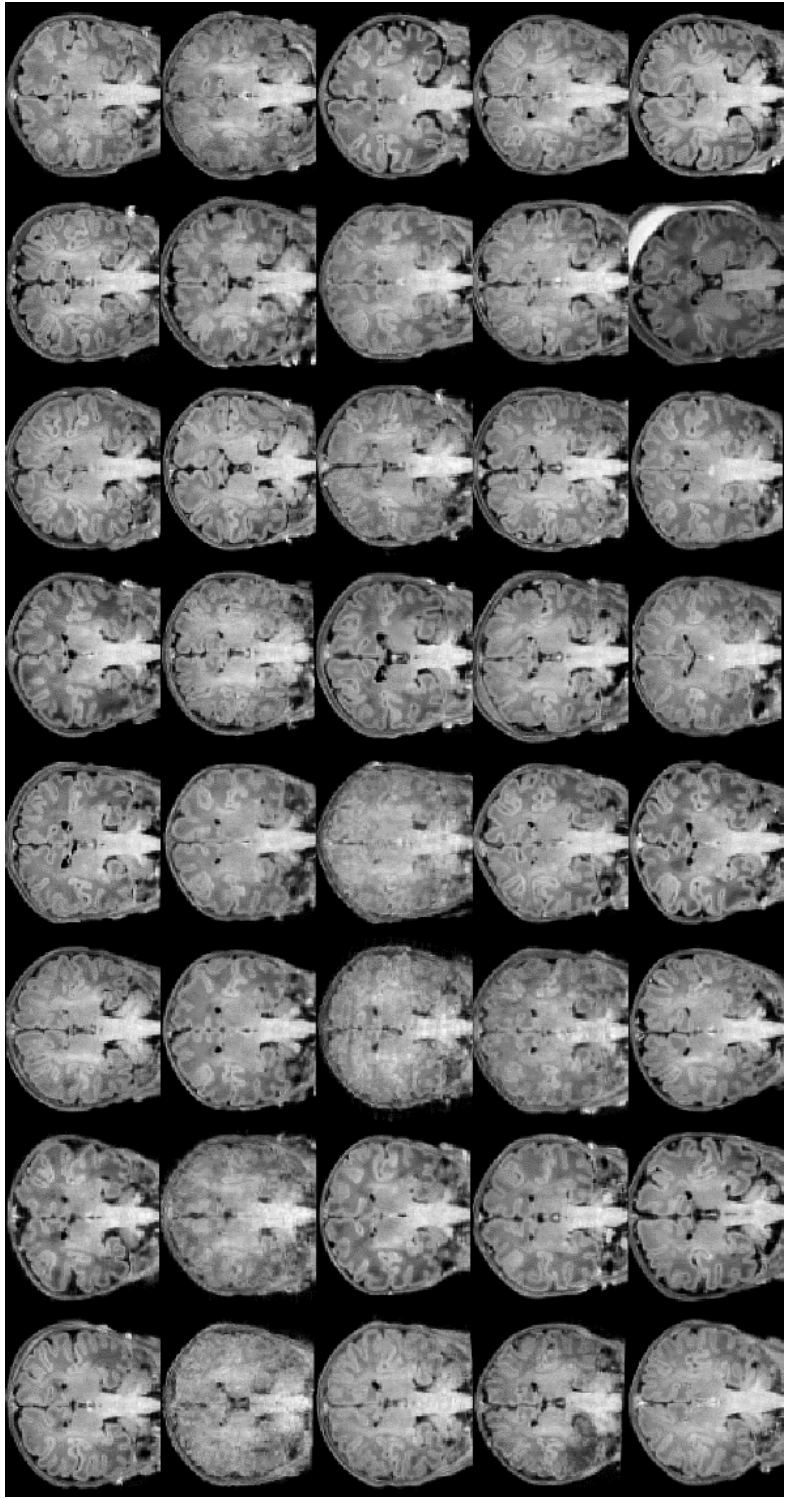

Figure 16: T1w input images of the first forty subjects constituting the recent data release of the "The Developing Human Connectome Project" dHCP project viewed in the coronal plane in an unbiased common affine coordinate system.

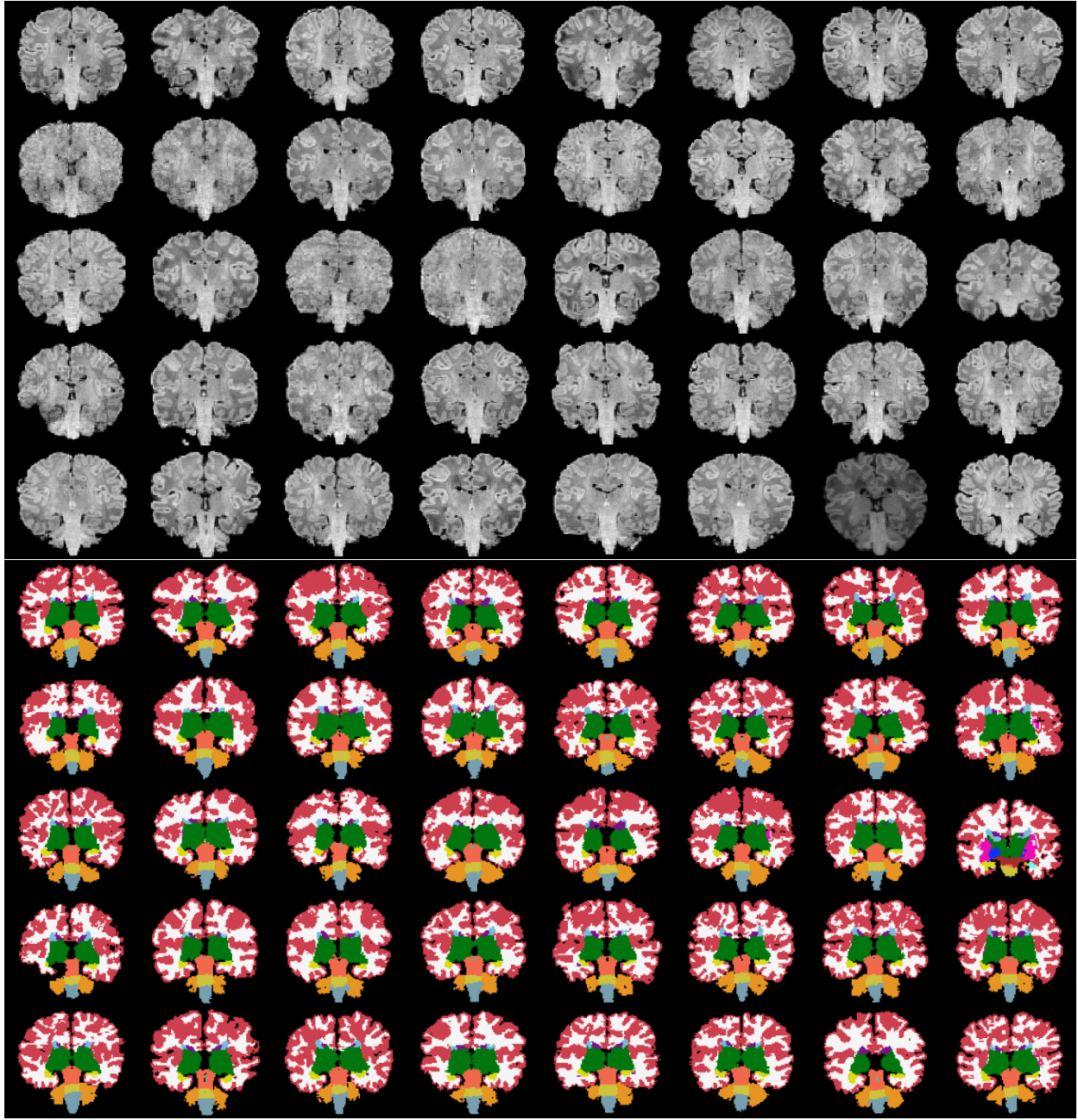

Figure 17: The first forty subjects constituting the recent data release of the "The Developing Human Connectome Project" dHCP project viewed in the coronal plane in an unbiased common affine coordinate system: (top) skull-stripped input images, and (bottom) segmented images using our new pipeline.

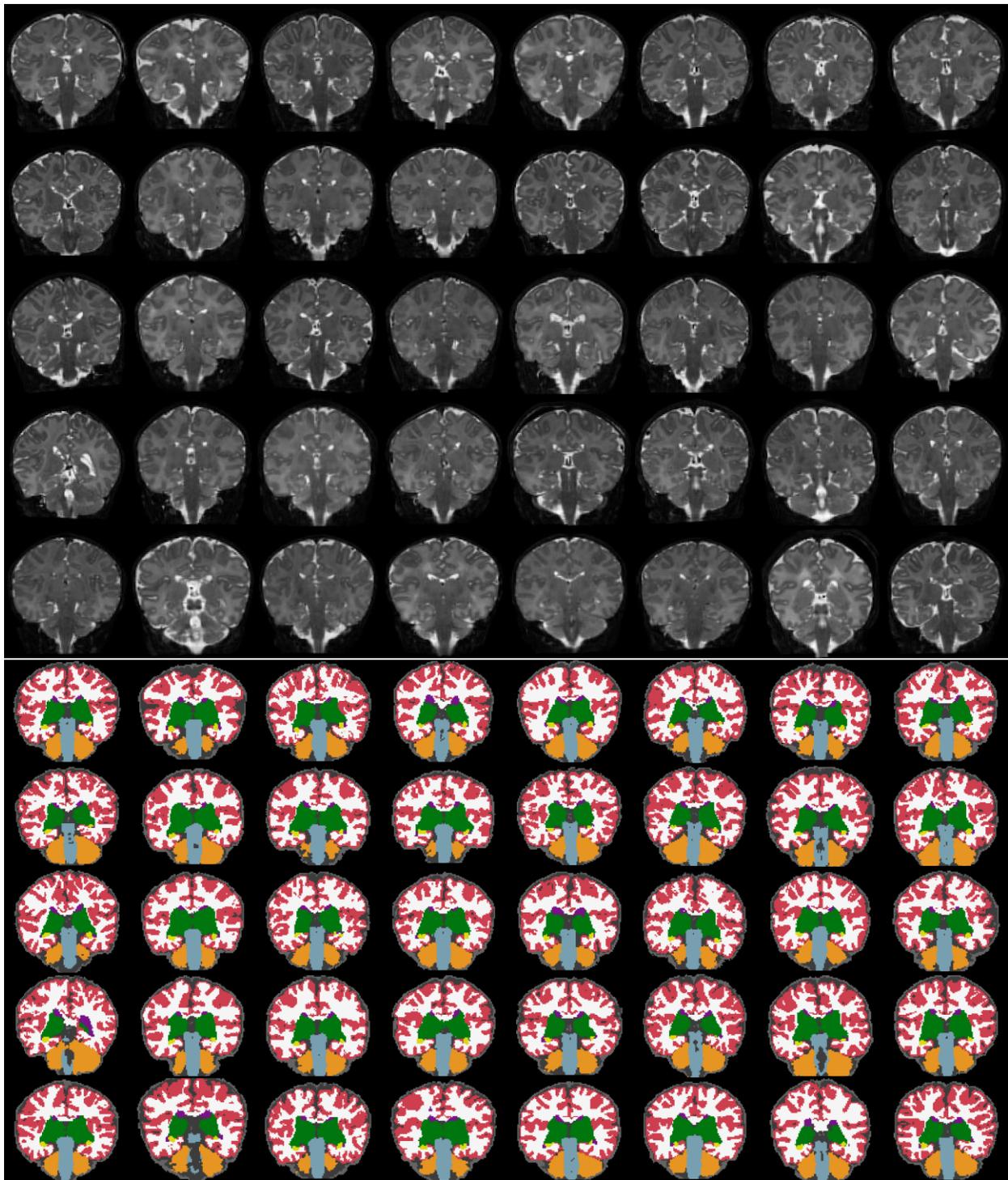

Figure 18: The first forty subjects constituting the recent data release of the "The Developing Human Connectome Project" dHCP project viewed in the coronal plane in an unbiased common affine coordinate system: (top) original T2w input images, and (bottom) dHCP released tissue segmentation outcomes (based on T2w images).

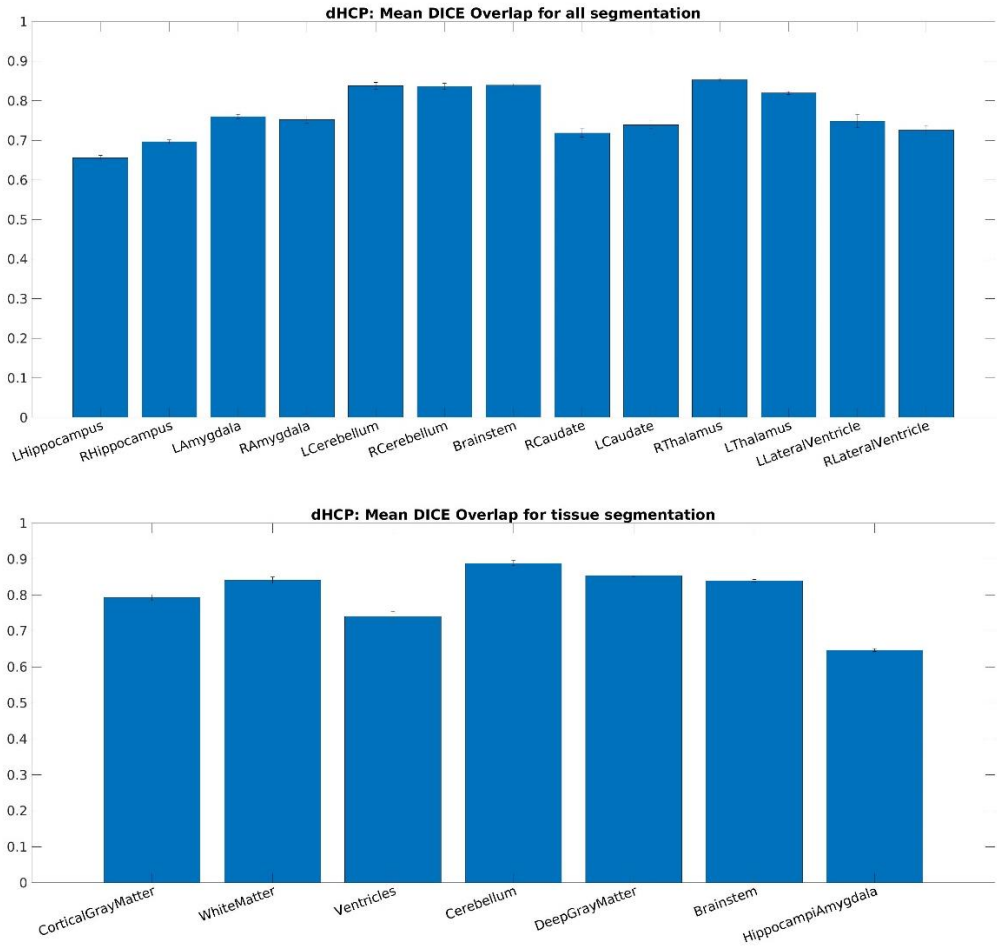

Figure 19: Mean Dice overlap measures computed on the DHCP dataset per segmentation labels: (top) "all" segmentation labels and (bottom) "tissue" segmentation labels released by the dHCP consortium.

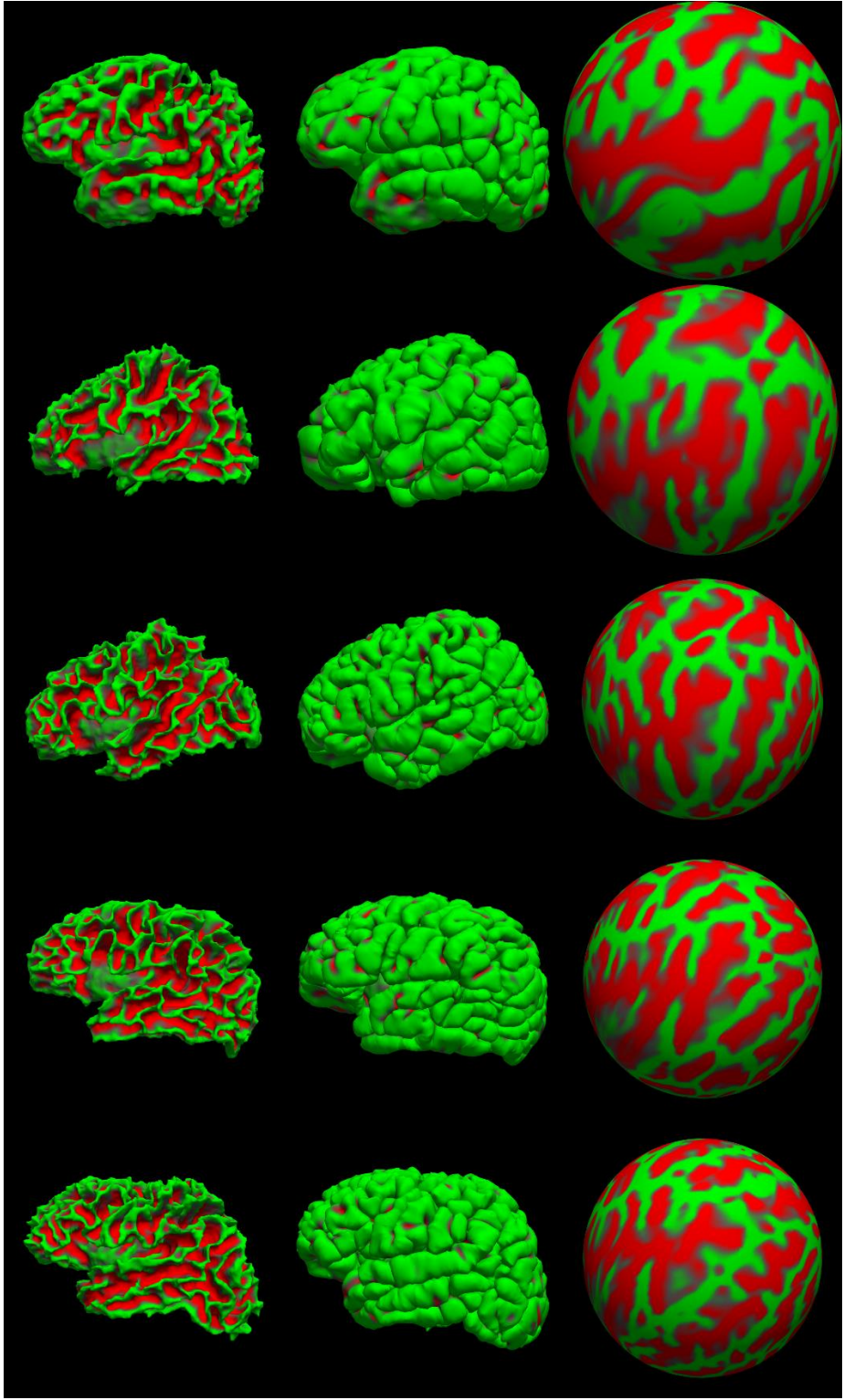

Figure 20: Surfaces generated for five sample subjects from our training dataset: (from top to bottom) newborn, 8mo, 12mo, 16mo, 18mo. From left to right: left hemisphere white surface, pial surface and spherical representation with a curvature map overlay.

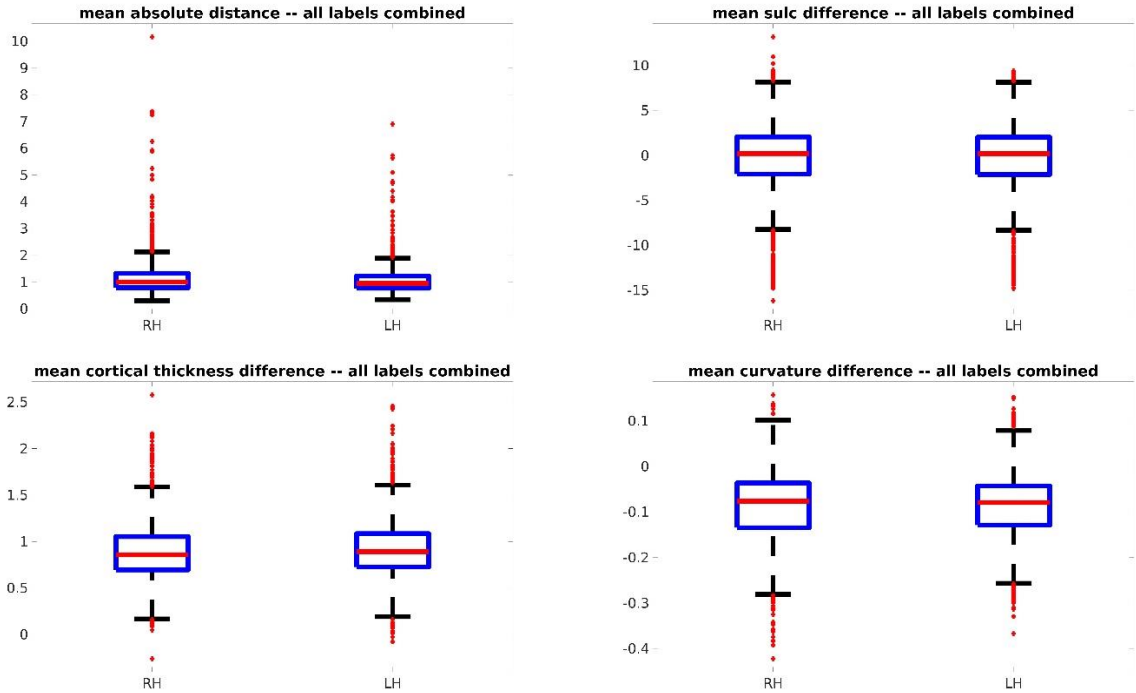

Figure 21: Boxplot displays comparing surface measures on the dHCP data set, per hemisphere, all labels combined: (top left) mean absolute distance, (top right) mean sulcal depth difference, (bottom left) mean cortical thickness difference and (bottom right) mean curvature differences.

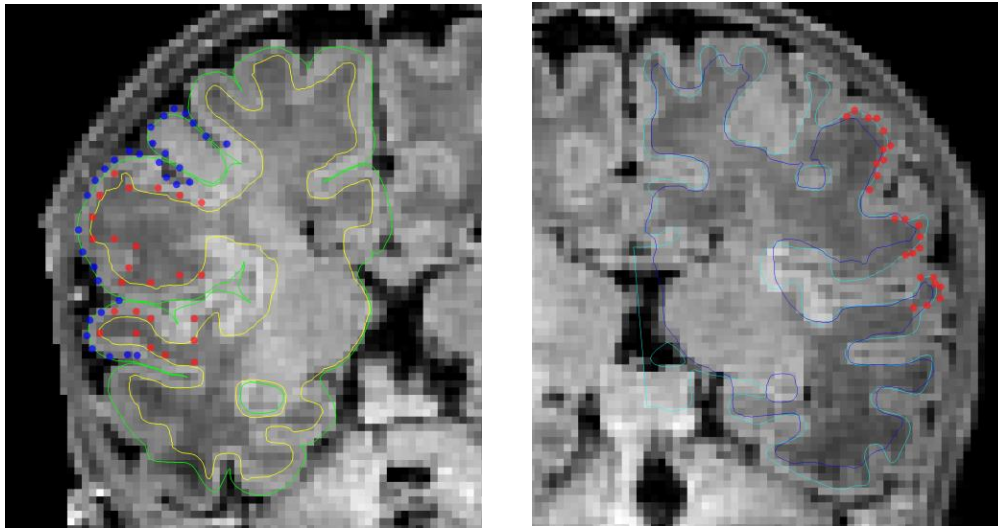

Figure 22: Examples of validation surface points placed on the right and left hemispheres of a randomly selected T1 weighted image from the *dHCP* data set, viewed on different coronal slices. (Left) validation surface points from both the pial (blue) and white (red) surfaces are shown along with our surface reconstruction solutions (light green – pial surface, yellow – white matter surface); (Right) validation surface points from the white (red) surface are shown along with the *dHCP* and our white matter surface reconstruction solutions (light blue – *dHCP*, dark blue – ours).

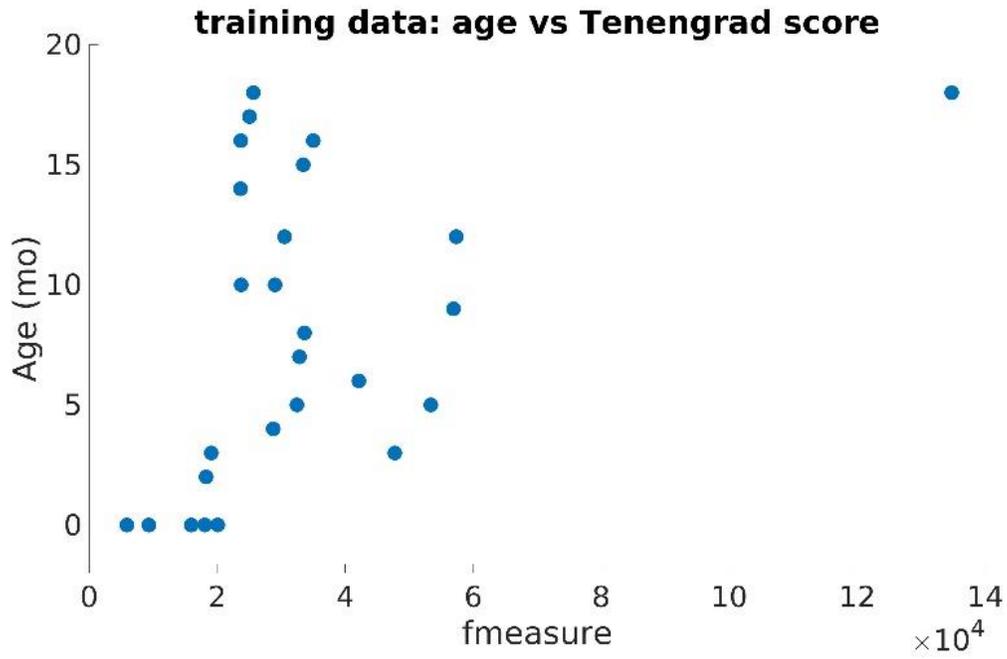

Figure 23: Age-at-scan vs the Tenengrad image sharpness metric (fmeasure) computed on the training dataset *(BCH_0-2 years)*.

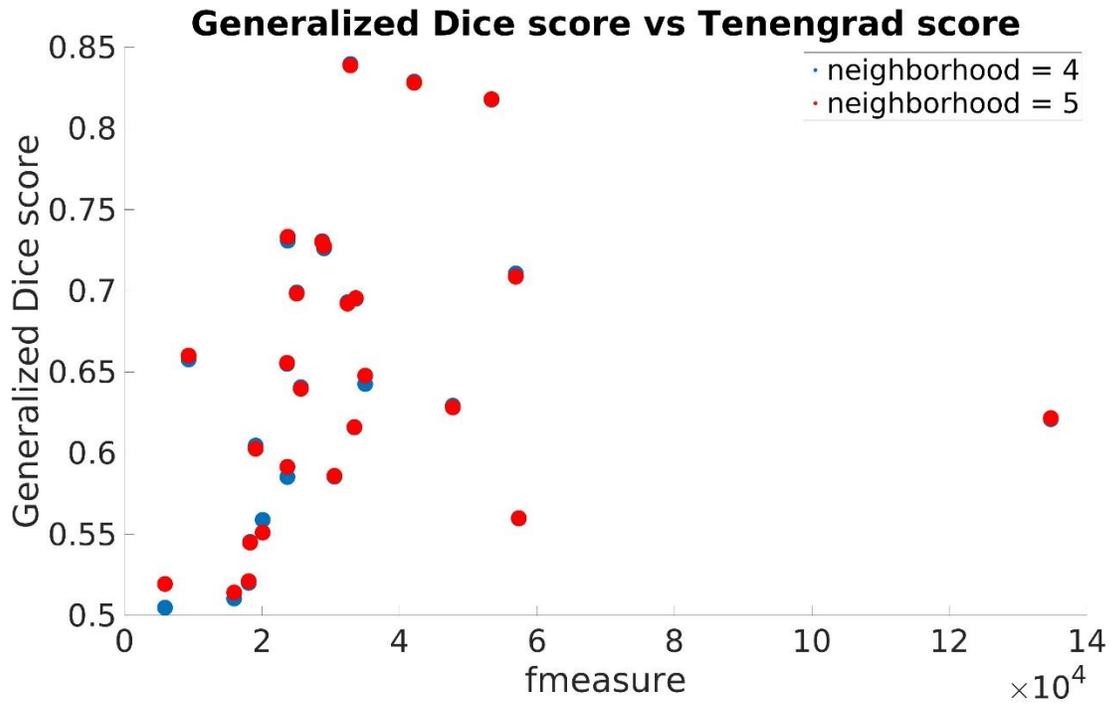

Figure 24: Generalized Dice score for each of training subjects using 4 and 5 as training neighborhood sizes vs the input image volumes' Tenengrad metric (fmeasure).

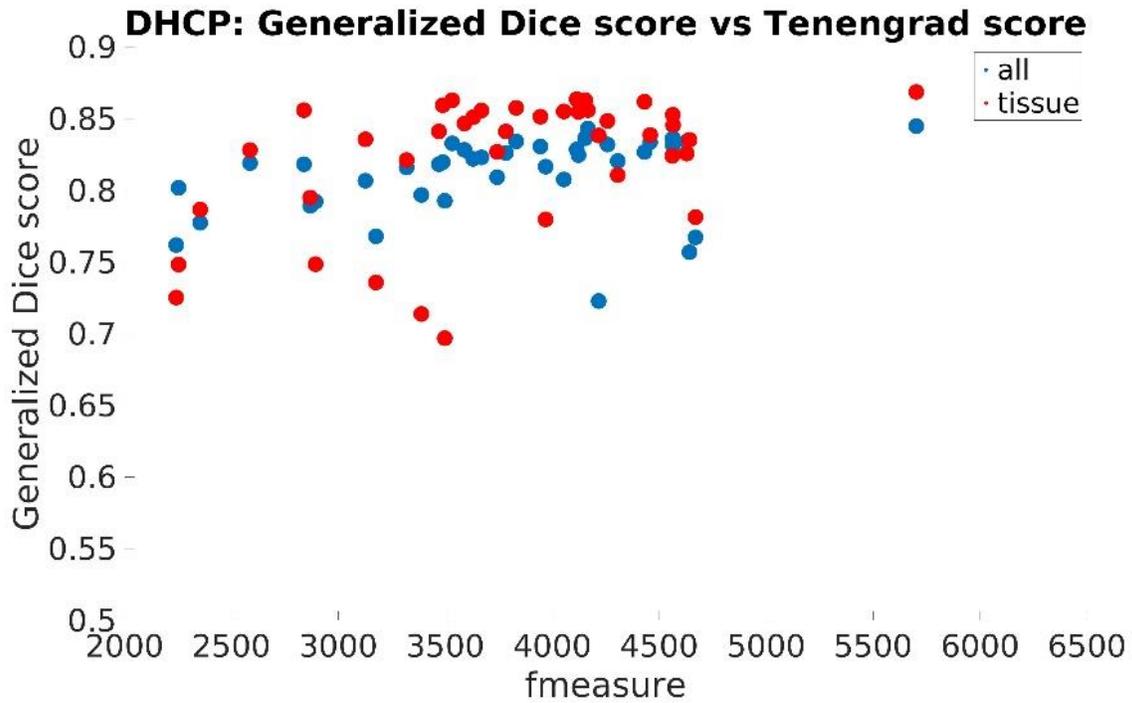

Figure 25: Generalized Dice score for each of dHCP subject using 5 as training neighborhood sizes vs the input image volumes' Tenengrad metric (fmeasure): "all" (in red) and common-with-our-pipeline "tissue" labels (in blue).

Table 1: Segmentation labels recovered by our proposed segmentation and their corresponding label IDs in FreeSurfer

| Label name | FS label | Label name | FS label |
|---|---|---|---|
| L/R CerebralWhiteMatter | (2,41) | 4th-Ventricle | (15) |
| L/R CerebralCortex | (3,42) | L/R Hippocampus | (17,53) |
| L/R LateralVentricle | (4,43) | L/R Amygdala | (18,54) |
| L/R CerebellarWhiteMatter | (7,46) | L/R Accumbens | (26,58) |
| L/R CerebellarCortex | (8,47) | L/R VentralDC | (28,60) |
| L/R Thalamus | (9,48) | Vermis | (172) |
| L/R Caudate | (11,50) | Midbrain | (173) |
| L/R Putamen | (12,51) | Pons | (174) |
| L/R Pallidum | (13,52) | Medulla | (175) |
| 3rd-Ventricle | (14) | | |

Table 2: Surface to label distances computed on two data sets (*BCH_0-2 years* and *dHCP* data set). Using 12 and 10 randomly selected subjects, respectively, shortest distances between points identified in the T1-weighted volume and the reconstructed surfaces were computed and the mean and standard deviation of the absolute value of these measurements are included in the Table.

|  |  | Infant FS | dHCP |
|---|---|---|---|
| *BCH_0-2 years* | **White matter surface** | 1.1732   (1.2525) | N/A |
|  | **Pial surface** | 0.9198   (0.9054) | N/A |
| *dHCP* | **White matter surface** | 1.1898   (1.1468) | 0.4585   (0.3384) |
|  | **Pial surface** | 0.8070   (0.8358) | 0.6470   (0.5205) |